\documentclass[12pt,journal,draftcls,letterpaper,onecolumn]{article}
\usepackage{wrapfig}
\usepackage{amsthm} 
\usepackage{comment} 
\usepackage{epsf}
\usepackage{graphicx}
\usepackage{subfigure}
\begin{document}
%\firstpage{1}

\title{Characterizing Discriminative Patterns}
\author{Gang Fang$^{1}\footnote{Corresponding author; Supplementary material: http://vk.cs.umn.edu/CDP}$, Wen Wang$^{1}$,
Benjamin Oately$^{1}$, \\
Brian Van Ness$^{2}$, Michael Steinbach$^{1}$, Vipin Kumar$^{1}$\\
$^{1}$Department of Computer Science, \\
$^{2}$Department of Genetics, Cell Biology, and Development\\
University of Minnesota, Minneapolis, MN 55455, USA.}
%\email{gangfang@cs.umn.edu}

\date{}

%\title[Construction and Functional Analysis of Human Genetic Interaction Networks with Genome-wide Association Data]{Construction and Functional Analysis of Human Genetic Interaction Networks with Genome-wide Association Data}
%\author[Fang et al.]{Gang Fang$^{1}\footnote{Corresponding author}$, Wen Wang$^{1}$, Vanja Paunic$^{1}$, Benjamin Oately$^{1}$, Majda Haznadar$^{2}$, Michael Steinbach$^{1}$, Brian Van Ness$^{2}$, Chad L. Myers$^{1}$, Vipin Kumar$^{1}$}
%\address{$^{1}$Department of Computer Science, $^{2}$Department of Genetics, Cell Biology, and Development\\
%University of Minnesota, Minneapolis, MN 55455, USA.}

\maketitle

\begin{abstract}

Discriminative patterns are association patterns that occur with disproportionate frequency in some classes versus others, and have been studied under names such as emerging patterns and contrast sets. Such patterns have demonstrated considerable value for classification and subgroup discovery, but a detailed understanding of the types of interactions among items in a discriminative pattern is lacking. To address this issue, we propose to categorize discriminative patterns according to four types of item interaction: (i) driver-passenger, (ii) coherent, (iii) independent additive and (iv) synergistic beyond independent additive. The coherent, additive, and synergistic patterns are of practical importance, with the latter two representing a gain in the discriminative power of a pattern over its subsets. Synergistic patterns are most restrictive, but perhaps the most interesting since they capture a cooperative effect that is more than the sum of the effects of the individual items in the pattern. For domains such as biomedical and genetic research, differentiating among these types of patterns is critical since each yields very different biological interpretations. For general domains, the characterization provides a novel view of the nature of the discriminative patterns in a dataset, which yields insights beyond those provided by current approaches that focus mostly on pattern-based classification and subgroup discovery. This paper presents a comprehensive discussion that defines these four pattern types and investigates their properties and their relationship to one another. In addition, these ideas are explored for a variety of datasets (ten UCI datasets, one gene expression dataset and two genetic-variation datasets). The results demonstrate the existence, characteristics and statistical significance of the different types of patterns. They also illustrate how pattern characterization can provide novel insights into discriminative pattern mining and the discriminative structure of different datasets.% Codes for pattern characterization and supplementary documents are available at http://vk.cs.umn.edu/CDP

%Supplementary material \url{http://vk.cs.umn.edu/humanGI}, \href{mailto:gangfang@cs.umn.edu}{gangfang@cs.umn.edu}
\end{abstract}

\section{Introduction}
\label{sec:intro}

For data sets with class labels, association patterns \cite{agrawal1994fam,book2005kumar} that occur with disproportionate frequency in some classes versus others, can be of considerable value. We will refer to them as discriminative patterns \cite{hongcheng2007icde,hongcheng2008icde,hongcheng2008kdd,fang2010subspace,fang2010tkde,petra2009jmlr} in this paper, although these patterns have also been investigated under various names, such as emerging patterns \cite{epfirst1999}, contrast sets \cite{cset2001} and supervised descriptive rules \cite{petra2009jmlr}. Discriminative patterns have been shown to be useful for improving the classification performance \cite{hongcheng2007icde,karypi2005tkde,module2007plosone} and for discovering sample subgroups \cite{petra2009jmlr,kraljnovak2009cset}. %\footnote{\small Discriminative patterns are itemsets for binary data, collections of variable-value pairs for categorical data, and sets of variable-range pairs for continuous data.}

The algorithms for finding discriminative patterns usually employ a measure for the discriminative power of a pattern. Such measures are generally defined as a function of the pattern's relative support\footnote{\small Note that, in this paper, unless specified, the support of a pattern in a class is relative to the number of transactions (instances) in that class, i.e. a ratio between 0 and 1.} in the two classes, and can be defined either simply as the ratio \cite{epfirst1999} or difference \cite{cset2001} of the two supports, or other variations, such as information gain \cite{hongcheng2007icde}, Gini index, or odds ratio~\cite{book2005kumar} etc.

\begin{figure}%[!t]
	\centering
		\includegraphics[width=0.6\textwidth]{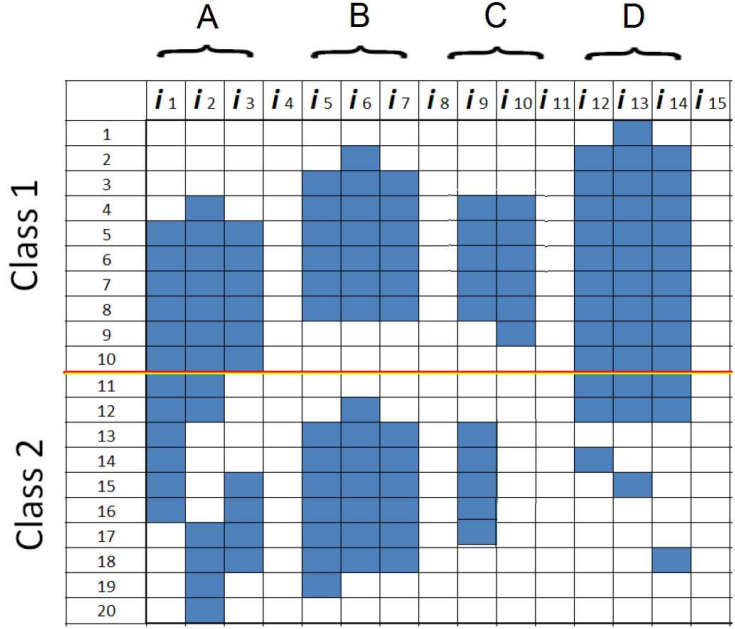}
	\caption{\small A sample data set with three discriminative patterns ($A$, $C$, $D$) and an uninteresting (non-discriminative) pattern ($B$)}
	\label{fig:toll}
\end{figure}

To introduce some key ideas about discriminative patterns and make the following discussion easier to follow, we use the measure that is defined as the difference of the supports ($\mathit{DiffSup}$) of an itemset in the two classes (originally proposed in \cite{cset2001} and used by its extensions~\cite{cset2007aime,kraljnovak2009cset}). Consider Figure~\ref{fig:toll}, which displays a sample dataset\footnote{\small The discussion in this paper assumes that the data is binary. Nominal categorical data can be converted to binary data without loss of information, while ordinal categorical data and continuous data can be binarized, although with some loss of magnitude and order information.} containing $15$ items (columns) and two classes, each with 10 instances (rows). In the figure, four patterns (sets of binary variables) can be observed: $A=\left\{i_1,i_2,i_3\right\}$, $B=\left\{i_5,i_6,i_7\right\}$, $C=\left\{i_9,i_{10}\right\}$ and $D=\left\{i_{12},i_{13},i_{14}\right\}$. $A$, $C$ and $D$ are discriminative patterns whose $\mathit{DiffSup}$ is $0.6$, $0.5$ and $0.7$ respectively. In contrast, $B$ is not discriminative with a relatively uniform occurrence across the classes ($\mathit{DiffSup} = 0$). %Furthermore, $D$ is a discriminative pattern whose individual items are also highly discriminative, while those of $A$ are not. Based on support in the whole dataset, $B$ is a frequent non-discriminative pattern, while $C$ is a relatively infrequent non-discriminative pattern. %Thus, in the pattern mining process, we want to prune patterns like $B$ and $C$ as early as possible, but keep potentially interesting patterns like $A$ and $D$. %\footnote{\small Described in the experimental setup.}

Although $A$, $C$ and $D$ are all considered to be discriminative because of their large $\mathit{DiffSup}$, several observations can be made about their different characteristics. First, one of the two items in $C$ has an individual $\mathit{DiffSup}$ of $0.6$ ($i_{10}$), while the other item ($i_9$) has a $\mathit{DiffSup}$ of $0$. Given that $C$ itself has a $\mathit{DiffSup}$ of $0.5$, it is obvious that the discriminative power of the pattern is mainly driven by $i_{10}$, while $i_{9}$ serves as a  ``passenger". Such driver-passenger effects result from the fact that measures for discriminative power such as $\mathit{DiffSup}$ only capture the joint discrimination of a pattern but ignore the specific contribution from the items in the pattern \footnote{Such driver-passenger patterns often result when a discriminating, low-support item is combined with a high-support, non-discriminating item. Similar issue exists in frequent pattern mining where a relatively low support item can form trivial patterns with many high-support items.}. Second, in contrast to $C$, the $\mathit{DiffSup}$ values of the three individual items in $A$ are $0$, $0.1$ and $0.2$, respectively, which are much lower than the $\mathit{DiffSup}$ of $A$ itself (0.6). This suggests that the items in $A$ have an incremental effect in their joint discriminative power. Third, in contrast to both $C$ and $A$, the three items in $D$ have $\mathit{DiffSup}$ values ($0.6$, $0.7$, $0.6$), which are very similar to that of the pattern itself ($0.7$). Thus, the three items in $D$, as well as their combination, show a coherent behavior in their ability to differentiate between class $1$ and class $2$.

%Understanding these different characteristics of discriminative patterns can be of significant value for domains such as biomedical and genetic research \cite{multigene2004nature,snpPair2007science, snpPair2007plos, snpPair2007naturegene}, where pattern interpretation is perhaps even more important than pattern discovery since interpretation can contribute to an understanding of the causes of disease from multiple genes or genetic variations such as single nucleotide polymorphisms (SNPs). Although classification models can play an important role in predicting disease phenotypes, they may not provide information about the mechanism of complex diseases like cancer. Thus, a detailed understanding of the interactions among genes can be of considerable value.

Patterns $A$, $C$ and $D$ have shown some of the characteristics of different types of \emph{interactions}\footnote{In this paper, we use \emph{interaction} to denote the relationship among the items in an itemset, and we use \emph{pattern} to denote the concept of itemset and used interchangably with \emph{itemset}.}. Indeed, some characteristics of such interactions have been discussed and studied. In particular, we can consider the discriminative power of a pattern as the confidence of an association rule by considering the class label as a special item. Then, the difference between the confidence of an association rule and the confidences of its subsets has been explored in the association rule mining community. Specifically, Bayardo et al. \cite{bayardo2000constraint} proposed a measure called \emph{improvement} as the difference between the confidence of an association rule (e.g. $Conf(X \rightarrow Y)$) and the maximal confidence of its simplifications (i.e. $max(\left\{Conf(X' \rightarrow Y)|X' \subset X\right\})$). The association rules that have positive improvement are called productive in \cite{webb2007discovering} and are considered to be more desirable than those rules with negative improvements. Similar approaches have also been proposed in the context of discriminative patterns. Garriga et al. \cite{petra2008jmlr} studied the closeness of discriminative patterns and proposed to remove a discriminative pattern (e.g. $X$ differentiates class 1 from class 2 by having higher support in class $1$ than in class $2$) if the support of $X$ is identical to any subset of $X$ in class $2$, because such patterns are guaranteed to have non-positive improvements.

\begin{figure}%[t!]
\centering
%\subfigure[\scriptsize Support ratio vs. Maximal-subset support ratio. \label{fig:mVSmax_rs}]{\includegraphics[width=.45\linewidth]{datauci_hepatic_plot1_101.eps}}
\subfigure[\scriptsize With support difference. \label{fig:mVSmax_ds}]{\includegraphics[width=.27\textwidth]{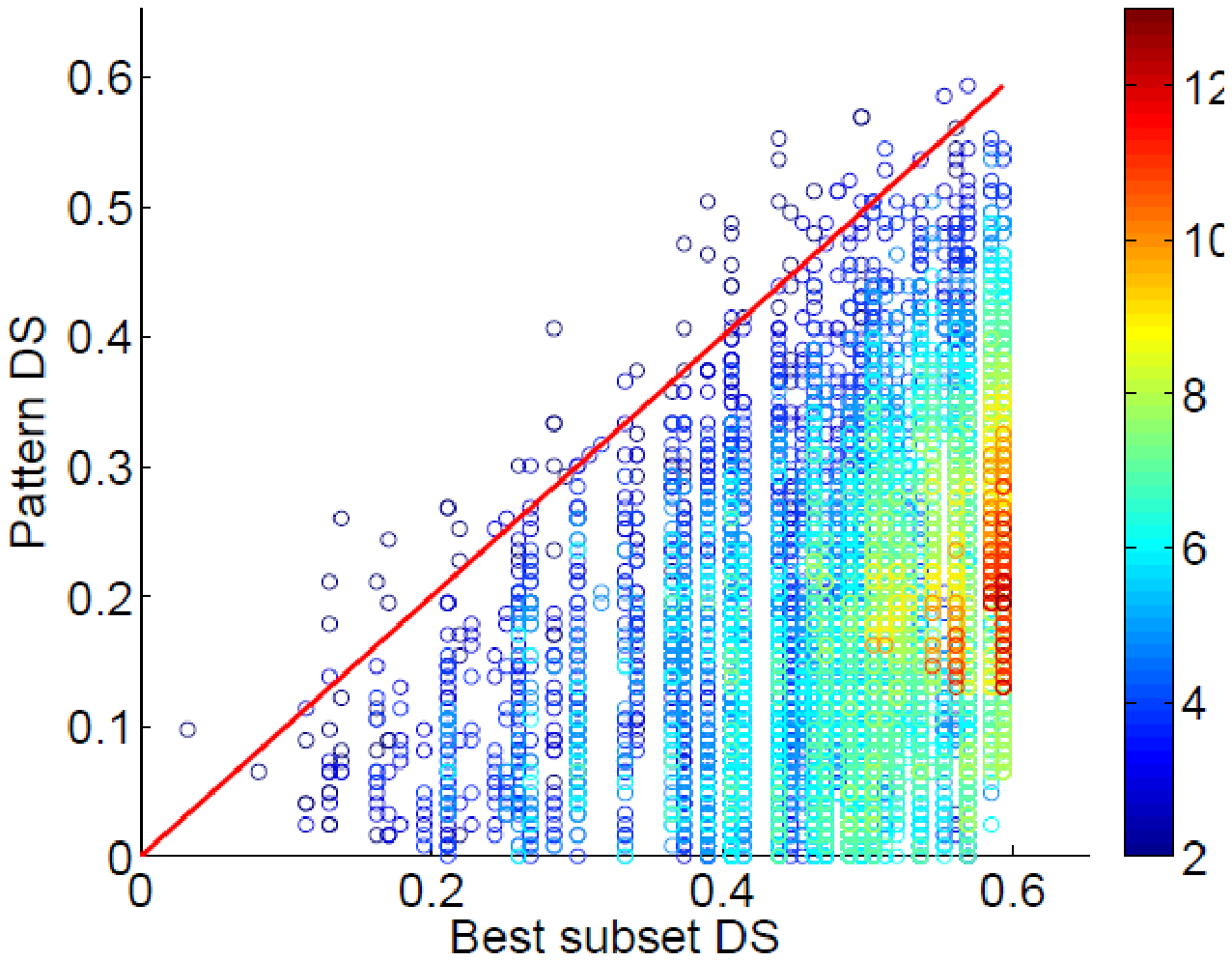}}
\subfigure[\scriptsize With $\chi^2$ statistic. \label{fig:mVSmax_chi2}]{\includegraphics[width=.27\textwidth]{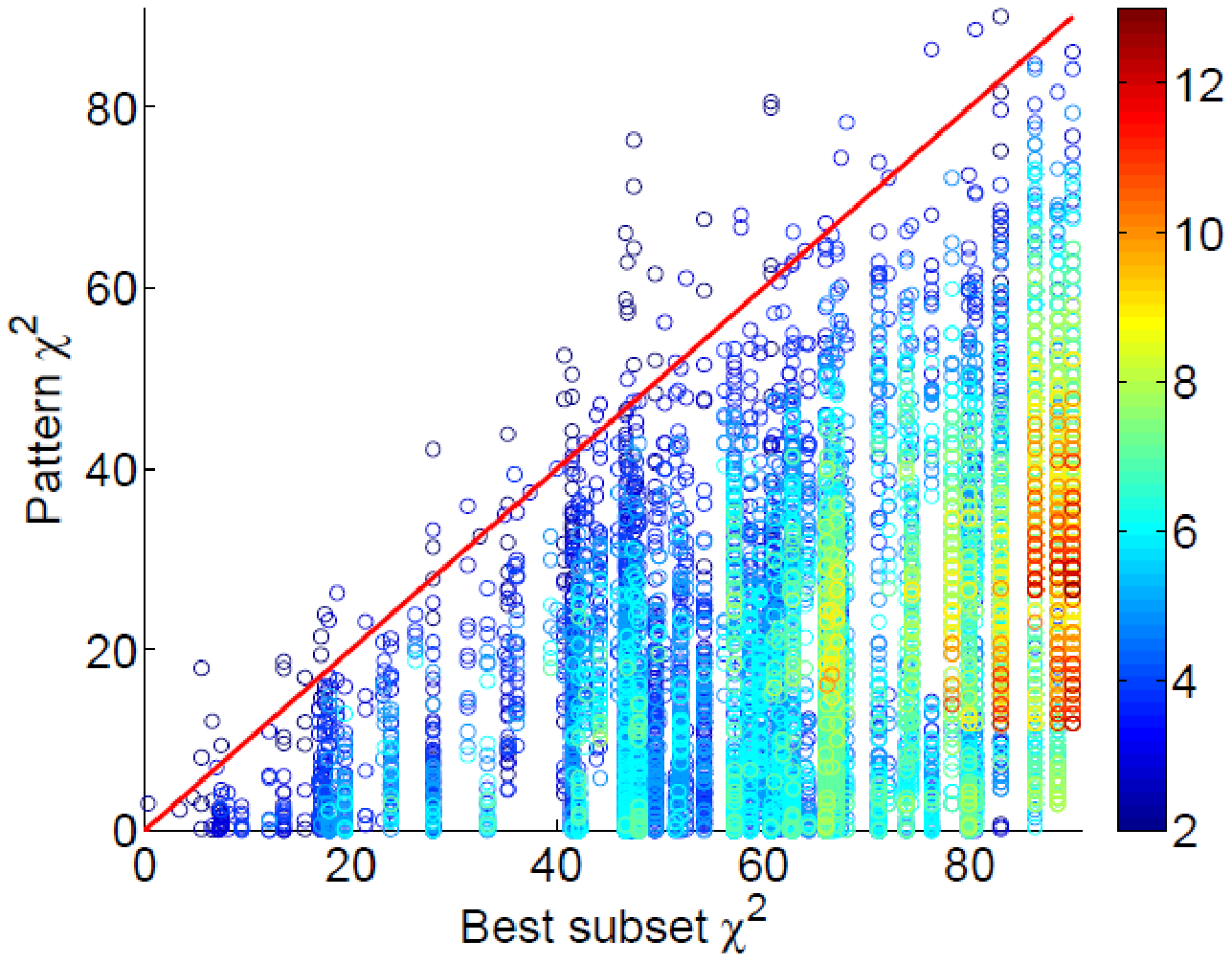}}
\subfigure[\scriptsize With mutual information. \label{fig:mVSmax_cmi}]{\includegraphics[width=.27\textwidth]{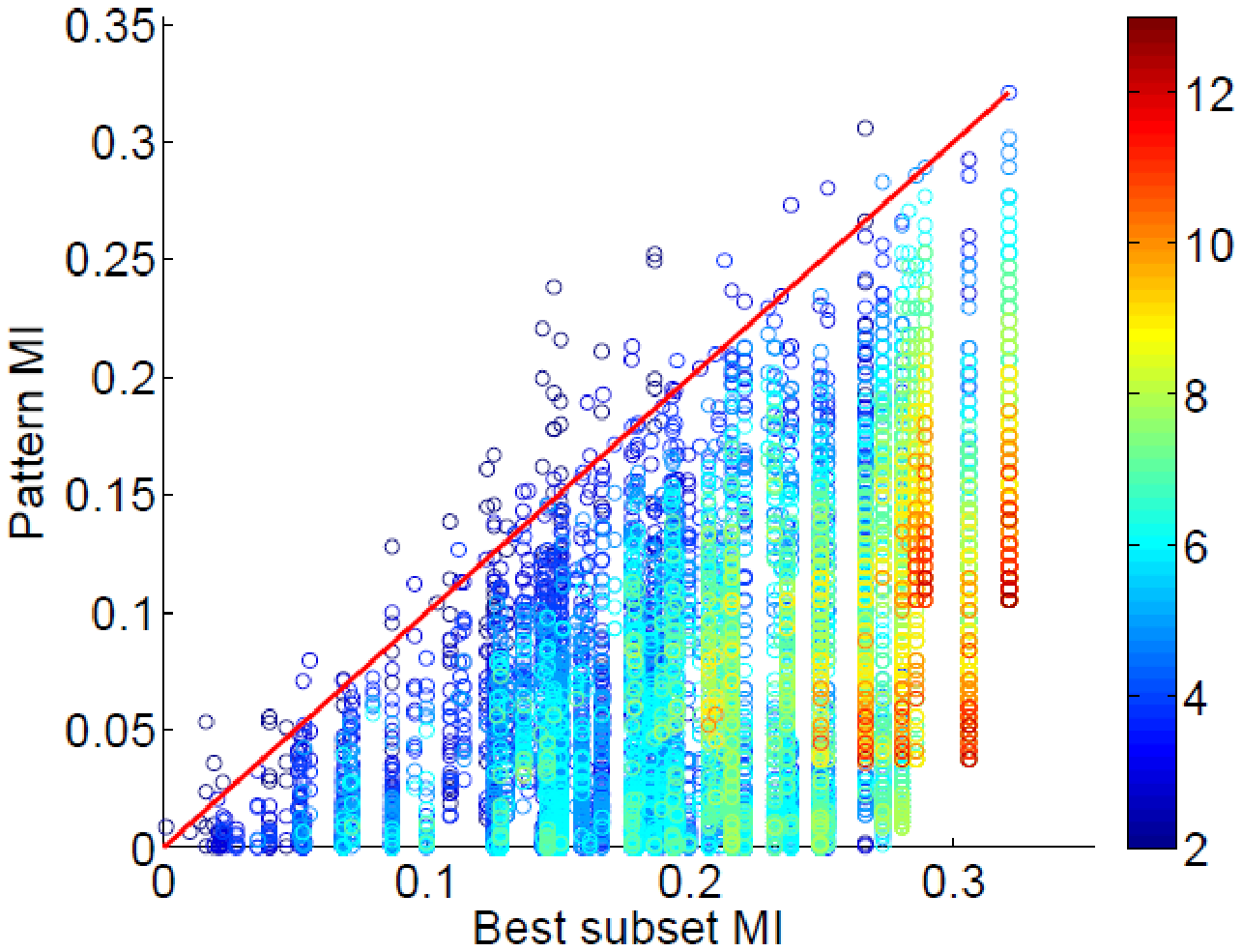}}
\caption{\small Comparing the discriminative power and maximal-subset discriminative power of a set of patterns discovered from the UCI Hepatic dataset, with three different measures for discriminative power. Each circle represent a pattern, with its color indicating pattern size (same for Figures \ref{fig:t2onAdult}, \ref{fig:t3t4tild}, \ref{fig:t3t4grid} and \ref{fig:t2t3t4uci}).}
\label{fig:mVSmax}
%\vspace{-0.4in}
\end{figure}

To illustrate the concept of \emph{improvement} and prepare for the following discussion, Figure \ref{fig:mVSmax} compares the discriminative power of a pattern with the best discriminative power of all its subsets, for all the frequent patterns ($minsup = 10\%$) in the Hepatic dataset (UCI \cite{uci2007}). Three measures are used in the subfigures (a), (b) and (c) respectively: support difference (\emph{DiffSup}), $\chi^2-$ statistic, and mutual information. The red line indicates $y=x$, which separates the patterns that have positive \emph{improvement} from those that have negative \emph{improvement}. A common observation consistent across the three subfigures is that, most patterns have at least one subset having higher discriminative power (negative improvement). In contrast, a small proportion of patterns have much higher discriminative power compared with their subsets (positive improvement). This contrast indicates that, although some combinations of items have a reasonably high joint association with a class variable, the actual amount of \emph{improvement} can vary greatly from pattern to pattern. 

As shown by existing studies \cite{bayardo2000constraint,webb2007discovering} as well as by Figure \ref{fig:mVSmax}, adding constraints on \emph{improvement} can reduce the number of interesting association and discriminative rules substantially. However, the current study of the different types of interactions in discriminative patterns is lacking in the following respects:

\begin{enumerate}

	\item The type of interaction captured by \emph{improvement} is only one of several interesting types of interactions. In other words, a discriminative pattern could have an interesting interaction even if it has a close-to-zero or even a negative \emph{improvement} value. For example, pattern $D$ shown in Figure \ref{fig:toll} does not have an improvement of discriminative power compared to that of its subsets, and may simply be due to the existence of multiple redundant discriminative features. However, such coherent differentiation of three items may still be interesting in certain domains. Specifically, in the field of differential gene-expression module discovery, a discriminative pattern like $D$ may indicate a functional module or protein complex. A specific example will be given in section \ref{sec:t2inter}.

	\item Even for the type of interactions captured by \emph{improvement}, a further understanding of the improvement in discriminative power is possible. For example, a large \emph{improvement} can either result from an independent additive aggregation of several items with separate (unrelated) association with a class variable, or a synergistic aggregation beyond the independent addition. Differentiating these different types of interactions (Section \ref{sec:intertypes}) can be useful for biomedical informatics because they generally lead to very different types of interpretation for a disease-genetic association \cite{costanzo2010genetic}. More generally, for other real-life applications, understanding different types of postive \emph{improvements} can help us understand the discriminative structure of a dataset.
		
	%\item Existing discriminative pattern mining algorithms mostly use measures for interaction as constraint to discover discriminative patterns, rather than consider the discriminative interaction itself as the objective function. Specifically, in the context of discriminative pattern mining, the patterns of most interest are those have top discriminative power and a reasonable interaction. In contrast, in the context of discriminative interaction minig, the patterns of best interest are those have have top interactions, even if their discriminative power are not as high as many others. From a statistical perspective, these two seemingly similar setup could lead to entirely different discoveries, after correcting for the multiple hypothesis testing, i.e. controlling the type I error rate or false discovery rate \cite{webb2007discovering,heikki2009tellme}.

\end{enumerate}

Aiming at a systematic understanding on the different types of interactions that are not captured by existing work, we motivate, formulate and design comprehensive experiments on the characterization of discriminative interactions from a general perspective of the discriminative pattern mining community. The major contributions of the paper are:

\begin{enumerate}
		\item	We categorize discriminative patterns into four groups based on the following types of interactions: (i) driver-passenger, (ii) coherent, (iii) independent additive and (iv) synergistic beyond independent addition.
	
	\item We present and discuss the properties and utility of the four interaction types we define. We also discuss the relationship of the four pattern types to one another. %Either of (ii), (iii) or (iv) constitutes a different, but potentially interesting type of discriminative pattern. Either (iii) or (iv) exhibits discriminative power beyond that of its subsets. The last type (iv) is the most restrictive yet potentially most interesting as it captures a purely cooperative effect beyond that of additive addition.
	
	%\item We formulate different null hypotheses for the discovery of each type of discriminative patterns, and motivate the necessity to correct multiple hypothesis testing for type I error, via a non-parametric randomization procedure. This statistical design provides a unbiased estimation of discriminative interactions. This framework, commonly used in the biomedical literature, has so far received little attention in the discriminative pattern mining community. Thus, we believe that it can also serve as a standard process to have a solid estimation of the statistical significance of discriminative pattern discovery.
	
	\item We design comprehensive experiments on various types of real datasets including ten UCI datasets, a gene expression dataset and two genetic variation (SNP) datasets. The results demonstrate the existence, characteristics and statistical significance of the different types of patterns. They also illustrate how pattern characterization can provide novel insights into discriminative pattern mining and the discriminative structure of different datasets. %For domains such as biomedical and genetic research, differentiating these different types of interactions is critical because they lead to different types of biological interpretations.
	
\end{enumerate}	

The rest of the paper is organized as follows. In Section \ref{sec:intertypes}, we discuss different types of interactions and define four types of discriminative patterns. In section \ref{sec:interresults}, we describe the datasets and experimental results. Related work on discriminative pattern mining is briefly summarized in Section \ref{sec:interrelatedwork}, followed by conclusions and future work in Section \ref{sec:interconclusion}.

\section{Different Types of Interactions and the Corresponding Groups of Discriminative Patterns}
\label{sec:intertypes}

In this section, we describe four types of item interactions and categorize discriminative patterns into four groups correspondingly. We also investigate their properties and their relationship to one another.

First we describe some terminologies that will be used through the rest of the section.

Let $D$ be a dataset with a set of items, $I = \left\{i_1,i_2,...,i_{|I|}\right\}$, two class labels $+$ and $-$, and a set of $n$ labeled instances (transactions), $D = \left\{(\textbf{\textit{x}}_i,y_i)\right\}_{i=1}^n$, where $\textbf{\textit{x}}_i \subseteq I$ is a set of items and $y_i \in \left\{+,-\right\}$ is the class label for $\textbf{\textit{x}}_i$. The two sets of instances that respectively belong to the class $+$ and $-$ are denoted by $D^{+}$ and $D^{-}$, and we have $|D| = |D^{+}| + |D^{-}|$. For an itemset $\alpha \subseteq I$, the set of instances in $D$, $D^{+}$ and $D^{-}$ that contain $\alpha$ are denoted by $D_{\alpha}$, $D_{\alpha}^{+}$ and $D_{\alpha}^{-}$ respectively. Let $p_{\alpha}$, $p_{\alpha}^{+}$ and $p_{\alpha}^{-}$ be support of $\alpha$ in $D$, $D^+$ and $D^-$ respectively, all relative to the entire set of transactions, i.e. $\frac{|D_{\alpha}|}{|D|}$, $\frac{|D_{\alpha}^{+}|}{|D|}$ and $\frac{|D_{\alpha}^{-}|}{|D|}$. Let $p^+$ and $p^-$ be $\frac{|D^{+}|}{|D|}$ and $\frac{|D^{-}|}{|D|}$ respectively.%. .

We use mutual information (MI) as  representative measure for discriminative power among many others such as the support ratio, support difference and $\chi^2$-statistic shown in section \ref{sec:intro}. This is because $MI$ is based on information theory, which makes one of the interaction measures to be presented later easy to interpret. The $MI$ between an itemset $\alpha$ and the class variable $C$ is computed as follows:

\begin{equation}
\scriptsize
%\vspace{-0.1in}
MI(\alpha;C) = \sum_{c \in \left\{+,-\right\}}\left(p_{\alpha}^{c}log\left(\frac{p_{\alpha}^{c}}{p_{\alpha}p^c}\right) + q_{\alpha}^{c}log\left(\frac{q_{\alpha}^{c}}{q_{\alpha}p^c}\right)\right),
\label{eq:mi}
\end{equation}

where $q_{\alpha}$, $q_{\alpha}^{+}$ and $q_{\alpha}^{-}$ are $1-p_{\alpha}$, $1-p_{\alpha}^{+}$ and $1-p_{\alpha}^{-}$ respectively. Note that, in this paper, $MI$ is always normalized by the entropy of the class variable ($H(C)$), after which, it ranges from $0$ to $1$.

\subsection{Driver-passenger Interaction (T1)}

Pattern $C$ shown in section \ref{sec:intro} is an illustration of discriminative patterns with a driver-passenger interaction, where the driver and the passenger are both a single item in the pattern. More generally, any discriminative pattern with a subset having similar discriminative power as the entire pattern while another disjoint subset in the pattern showing weak discriminative power are considered to have a \emph{driver-passenger} interaction. Formally, we define the discriminative patterns with this type of interaction (T1) as follows:

%\newdef{definition}{Definition}
%\begin{definition}
%\label{def:t1inter}
\textbf{Definition 1}: An itemset $\alpha$ is a T1 discriminative pattern if the following criteria are met together for $\delta > 0$, $j > 0$, $\epsilon > 0$:
\begin{equation}\label{equ:t1inter}
\centering
\begin{array}{ll}
(a) & MI(\alpha,C) > \delta,\\
(b) & \exists \alpha' \subset \alpha, |MI(\alpha,C)-MI(\alpha',C)| < j, \\
(c) & \exists \alpha'' \subseteq (\alpha - \alpha'), MI(\alpha'',C) < \epsilon.
\end{array}
\end{equation}
%\end{definition}

Criterion (a) is a general requirement of the discriminative power of an itemset, which will also be used in the definition of the other types of discriminative patterns. Criteria (b) and (c) require the existence of at least one driver and at least one passenger in $\alpha$, respectively. Similar to $C$ in Figure \ref{fig:toll}, $T1$ discriminative patterns are generally not interesting because the passengers are included in a pattern as a purely mathematical consequence rather than an interpretable relationship with the other items in the pattern. Thus, in the rest of the paper, we will focus on the other three types of interactions that can serve as evidence of meaningful relationship among the items in a pattern.

\subsection{Coherent Interaction (T2)}
\label{sec:t2inter}

The illustrative pattern $D$ in Figure \ref{fig:toll} represents a type of interaction in which every item in a pattern is contributing with a discriminative power similar to that of the entire pattern. We call this a \emph{coherent interaction}, and refer to patterns having this type of interaction T2 patterns.

\textbf{Definition 2}: An itemset $\alpha$ is a T2 discriminative pattern if the following criteria are met together for $\delta > 0$, $j > 0$:
\begin{equation}\label{equ:t2inter}
\centering
\begin{array}{ll}
(a) & MI(\alpha,C) > \delta,\\
(b) & incoherence(\alpha) <  j,\\
(c) & \forall i \in \alpha, direction(i) = direction(\alpha).
\end{array}
\end{equation}
%\end{definition}

\begin{figure}[t!]
\centering
\subfigure[\scriptsize T2 patterns (upper-left). \label{fig:t2onAdult}]{\includegraphics[width=.56\textwidth]{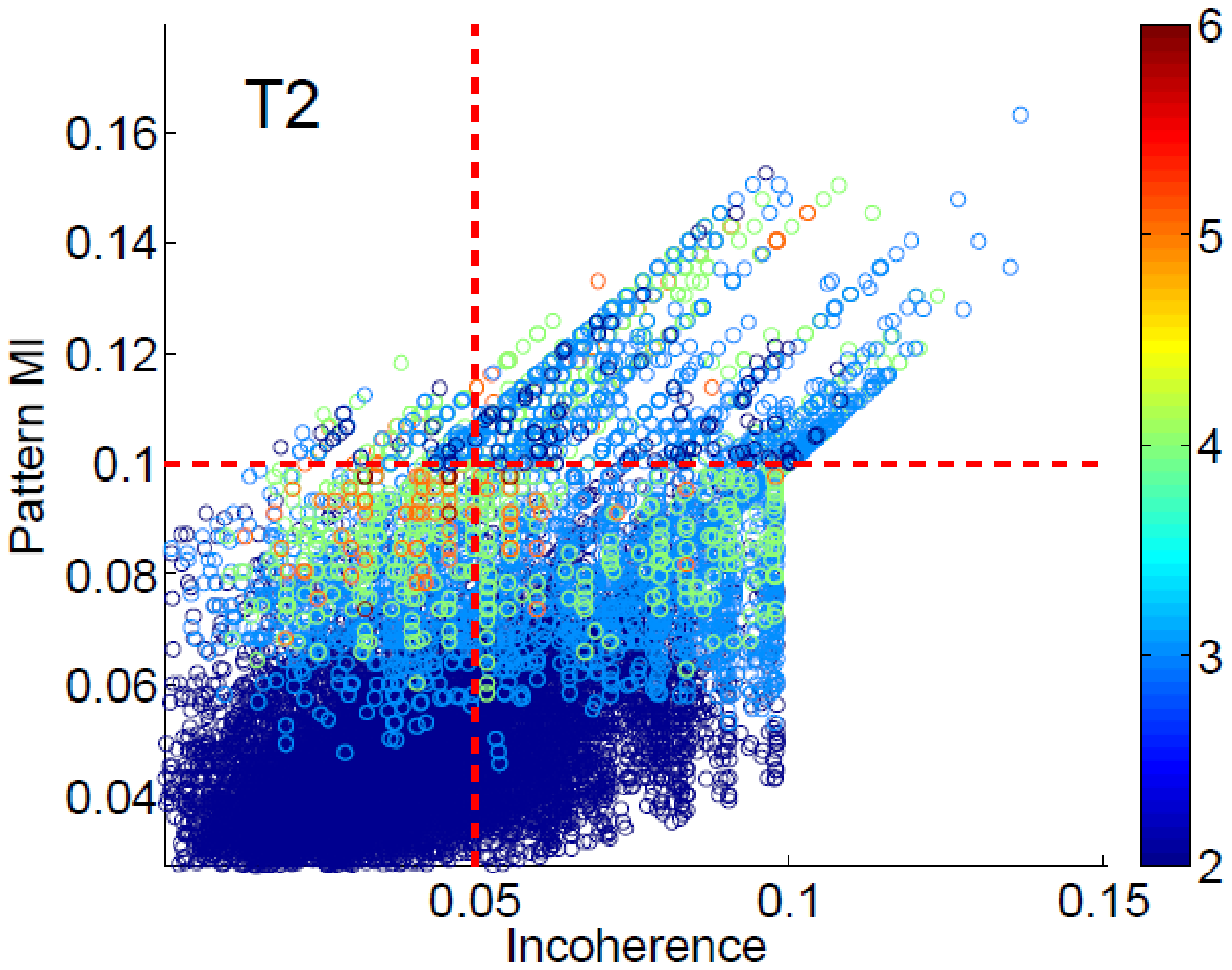}}
\subfigure[\scriptsize A T2 example. \label{fig:gustavoPatt}]{\includegraphics[width=.42\textwidth]{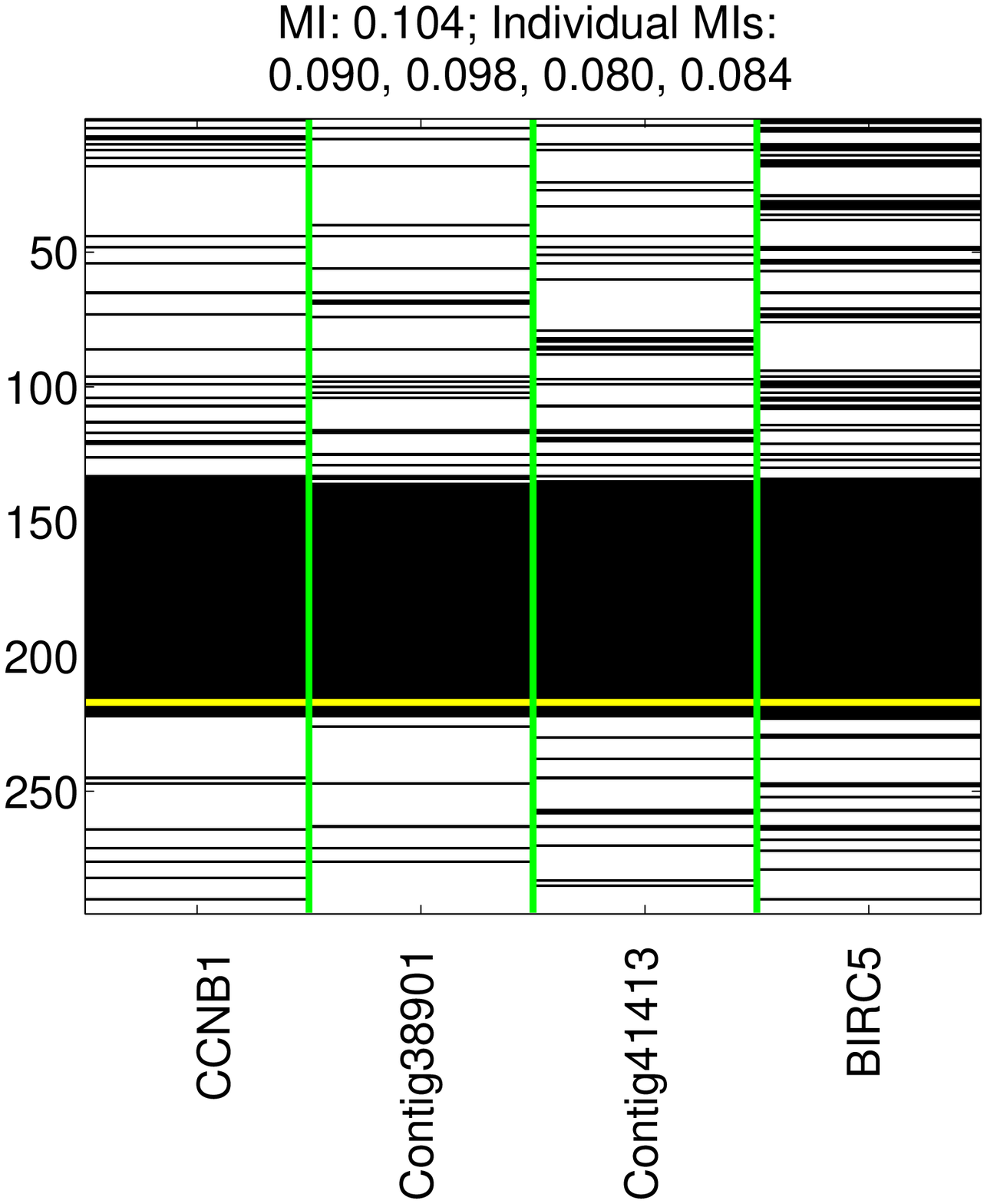}}
\caption{\small Illustration of T2 discriminative patterns on the gene expression dataset (described in section \ref{sec:interresults}). (a) The entire set of discovered patterns; (b) visulization of the pattern in a binary matrix format (black indicating 1's and white representing 0's similar as Fig. \ref{fig:toll}) with the horizontal yellow line separating the two classes and the vertical green lines separating genes from each other.}
\label{fig:t2inter}
%\vspace{-0.4in}
\end{figure}

The \emph{incoherence} in criterion (b) is calculated as the range \footnote{Difference between the maximal and the minimal value.} of values in $\left\{MI(\alpha,C)\right\} \cup \left\{MI(i,C)|i \in \alpha\right\}$. Criteria (a) and (b) capture the unique property of this type of coherent interaction, i.e. each individual item in a pattern has similar (controlled by $j$) discriminative power as the pattern itself. Given that $MI$ does not indicate the direction of the differentiation (i.e. a pattern or an item can be either more frequent in class $+$ or more frequent in class $-$), criterion (c) is further used to make sure that all the items in a pattern have the same differentiating directionality as the pattern itself. Figure \ref{fig:t2onAdult} illustrate the existence of T2 discriminative patterns with a real gene expression dataset. Each circle represents a pattern. The circles above the horizontal line meet criterion (a), and the circles on the left of the vertical line meet the criterion (b). Criterion (c) is implicitly enforced in the generation of the figure. The circles in the upper-left corner are $T2$ discriminative patterns. Note that the definition of different type of interactions is with respect to the specified parameters (here $\delta = 0.1$ and $j = 0.05$), rather than a clear-cut separation. With different parameter values, different set of patterns will be considered to have a certain interaction.

%Such correlation may either just due to the existence of multiple discriminative features that are redundant with each other (uninteresting), or may correspond to a functional module or protein complex that is associated with a disease in the context of differential gene-module discovery %In contrast, if a dataset has few or no $T2$ discriminative patterns, the discriminative features in the dataset (if exist) are expected to be uncorrelated with each other.

The essential difference between T1 and T2 discriminative patterns is that T1 patterns include passengers (guaranteed by the criterion (c) in Definition 1), while T2 patterns do not include passengers (guaranteed by the criterion (b) in Definition 2). This difference is what distinguishes T1, an uninteresting type of discriminative pattern, from T2, a a potentially interesting type of discriminative pattern. Specifically, if a dataset has many $T2$ discriminative patterns, we can speculate that it contains features that are discriminative and correlated with each other. Such correlation may either be due to the existence of multiple discriminative features that are redundant with each other (uninteresting), or may correspond to a functional module or protein complex that is associated with a disease in the context of differential gene-module discovery. For instance, figure \ref{fig:gustavoPatt} illustrates a T2 pattern discovered from the gene expression dataset of a study on breast cancer \cite{veer02} (section \ref{sec:datasets}). The genes in the pattern ($MI(\alpha,C) = 0.10$ and $incoherence(\alpha)  = 0.02$) demonstrate similar type of differentiating effect as $C$. Discovering such patterns rather than the individual items separately could provide valuable insights towards the understanding of gene interactions in complex diseases. Indeed, three genes in the pattern ($BIRC5$, $Contig38901$ and $Contig41413$) have been associated with breast cancer specifically\footnote{www.genecards.org}, and the other one ($CCNB1$) was identified as a general tumor-related gene \cite{park2007nf_ccnb1}. These facts suggest that the genes in the pattern may correspond to a functional module or protein complex.%Note that, this is another example that classification models themselves, although can aid the prediction of disease-phenotypes, are not as informative as discriminative pattern mining in term of assisting the understanding of the causes of a disease (critical for disease treatment)

\subsection{Independent-Additive Interaction and Synergistic Interaction beyond Independent Addition (T3 and T4)}
\label{sec:t3t4now}

In addition to coherent interaction, another type of interesting interaction in biomedical and genetic domains is a pattern containing a set of items (e.g. genes) that has better discriminative power than any of its subsets. Pattern $A$ illustrates such an example, i.e. the three individual items are not discriminative by themselves while they have a 100\% prediction confidence as a combination. %Such high interaction may serve as a valuable information that can be traced towards a detailed understanding of a disease mechanism. Not just for biomedical, such equally important for many other domains.%Such a increment of discriminativ power may provide insights about the interesting interactions among several items not only for the biomedical domain but for general fields as well.

As discussed in section \ref{sec:intro}, this type of interaction can be captured by existing measures such as \emph{improvement}, which is defined to be the difference between the discriminative power (e.g. MI) between a pattern and its best subset. However, a deeper understanding of the characteristics of the improvement in discriminative power is possible. For example, for pattern $A$ in Figure \ref{fig:toll}, the large \emph{improvement} can either result from an independent additive aggregation of several items with separate (unrelated) association, or a synergistic aggregation beyond the independent addition. Differentiating these different types of interactions is important because they generally lead to very different types of interpretation of a disease association. %More generally, for other real-life datasets, understanding different types of discriminative interactions can help us understand the complex interaction structure of a dataset.

Next, we will first discuss two different types of improvement interactions and then define another two types of discriminative patterns accordingly.

\subsubsection{Differentiating two types of improvement interactions}

Bayardo et al. \cite{bayardo2000constraint} defined \emph{improvement} in the context of association rule mining based on the \emph{confidence} of a rule. We first rewrite the \emph{improvement} ($Imp$) in the context of discriminative pattern mining based on MI as below:

\begin{equation}
%\scriptsize
%\vspace{-0.1in}
Imp^C(\alpha) = MI(\alpha,C) - max_{\alpha' \subset \alpha}(MI(\alpha',C)).
\label{eq:imp_cmi}
\end{equation}

To ease the motivation of different types of \emph{improvement}, we consider the following equation for a pair of items $\alpha = \left\{i_a,i_b\right\}$.

\begin{equation}
%\scriptsize
%\vspace{-0.1in}
Imp^C(\alpha) = MI(\alpha,C) - max(MI(i_a,C),MI(i_b,C)),
\label{eq:imp_cmi_pair}
\end{equation}

which is essentially the amount of additional information about the class variable $C$ that can be provided by the two items as a combination, compared to the information that each item can provide (the bigger one). This additional amount of information can either result from an independent additive aggregation of several items with separate (unrelated) association, or a synergistic aggregation beyond the independent addition. 

D. Anastassiou \cite{anastassiou2007computational} applied a measure called \emph{synergy} (originally used in neuroscience literature \cite{gawne1993independent}) to discover gene-gene interactions that are beyond the independent addition of all possible partitions of its subsets. In this paper, we leverage it to characterize discriminative patterns from a more general perspective.

We start from the following equation for calculating the synergy computation between a size-$2$ pattern $\alpha = \left\{i_a,i_b\right\}$ and a class variable $C$,

\begin{equation}
%\scriptsize
%\vspace{-0.1in}
Syn^C(\alpha) = MI(\alpha,C) - (MI(i_a,C) + MI(i_b,C)),
\label{eq:syn_pair}
\end{equation}

which is calculated as the amount of additional information about the class variable $C$ that can be provided by the two items as a combination, compared to the information that each of the item can provide independently (sum of the two individual $MIs$). Compared to Equation \ref{eq:imp_cmi_pair}, the essential difference between \emph{improvement} and \emph{synergy} is that, \emph{improvement} is with respect to the bigger $MI$ of the two, while \emph{synergy} is compared to the summation of $MIs$ of the two. Indeed, the summation of the mutual information of two items is used in information theory to represent the combined effect of two items with independent association with a class variable \cite{anastassiou2007computational}. Thus, \emph{synergy} can be leveraged to refine the discriminative patterns with positive \emph{improvement}, based on the characteristics of an improvement.

\begin{figure}%[!t]
	\centering
		\includegraphics[width=0.45\textwidth]{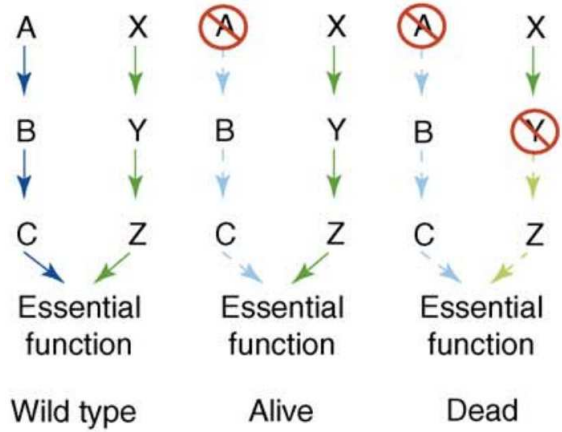}
	\caption{\small Illustration of the mechanism of synergistic interaction in the context of yeast genetic interaction (Figure taken from Costanzo et al \protect\cite{costanzo2006experimental}).}
	\label{fig:gi_motivation_costanzo}
\end{figure}

In order to provide an intuitive understanding, Figure \ref{fig:gi_motivation_costanzo} illustrates an underlying mechanism of synergistic interaction in the context of yeast genetic interaction. Two distinct pathways are shown in the figure, i.e. $A\rightarrow B\rightarrow C$ and $X\rightarrow Y\rightarrow Z$, which impinge on a common biological process that is essential to the survival of a yeast cell (the wild type). Due to parallel structure, the two pathways can compensate for the loss of the other, and thus a genetic perturbation (natural variations) on either of the two pathways separately (e.g. perturbation only in $A$) fails to cause any observable defects in cell survival. However, the simultaneous perturbations in $A$ and $Y$ disrupt both pathways and result in the lethality the cell. In this example, $A$ and $Y$ have a synergistic interaction with respect to the class label (survival or not) of a cell.

Equation \ref{eq:synergy_general} gives the general definition of \emph{synergy} for an itemset $\alpha$ beyond pairs (also defined in \cite{anastassiou2007computational}).

\begin{equation}
\scriptsize
%\vspace{-0.1in}
Syn^C(\alpha) = MI(\alpha,C) - max_{all\ partitions\ into\ \left\{S_i\right\}}\sum_iMI(S_i,C),
\label{eq:synergy_general}
\end{equation}

where a partition is defined as a collection $\left\{S_i\right\}$ of disjoint subsets $S_i$ whose union is $\alpha$. For example, for a size-$3$ pattern($\alpha = \left\{i_a,i_b,i_c\right\}$),

\begin{equation}
\scriptsize
%\vspace{-0.1in}
Syn^C(\alpha) = MI(\alpha,C) - max\left\{\begin{array}{l}
    MI(i_a,C)+MI(i_b,C)+MI(i_c,C)\\
    MI(i_b,C)+MI(\left\{i_a,i_c\right\},C)\\
    MI(i_c,C)+MI(\left\{i_a,i_b\right\},C)\\
    MI(i_a,C)+MI(\left\{i_b,i_c\right\},C)\\
  \end{array} \right.
\label{eq:synergy_triplet}
\end{equation}

This generalized definition is consistent with the intuition that \emph{synergy} is the additional amount of information about a class variable provided by an integrated discriminative power compared with what can be best achieved after breaking the pattern into components by the sum of the contributions of these components. The partition of the set of factors that is chosen in this formula is the one that maximizes the sum of the amounts of mutual information connecting the subsets in that partition with the class variable, and we will refer to it as the \emph{best aggregated MI}. Note that, the computational complexity of \emph{synergy} for an itemset of size $n$ is $O(B(n))$, where $B(n)$ is the $n^{th}$ Bell number\footnote{en.wikipedia.org}, which increases in a dramatically fast speed. In practice, to avoid unnecessary computations, we actually only need to compute \emph{synergy} (as well as \emph{best aggregated MI}) for those patterns with positive \emph{improvement}, which is much more efficient to compute, i.e $O(n)$. %This applies to Figures \ref{fig:t4inter_pre}, \ref{fig:t4inter_def}, \ref{fig:survivalt4} and \ref{fig:sonart4}.

%Note that because we always normalize $MI$ by $H(C)$, $Syn_C(S_i,S_j)$ also ranges from $-1$ to $+1$. Say $MI S_i,S_j$ in information theory Show what's positive and what's negative.

\begin{figure}[t!]
\centering
\subfigure[\scriptsize Positive $Imp$ (above $y=x$) \label{fig:t3inter_pre}]{\includegraphics[width=.49\textwidth]{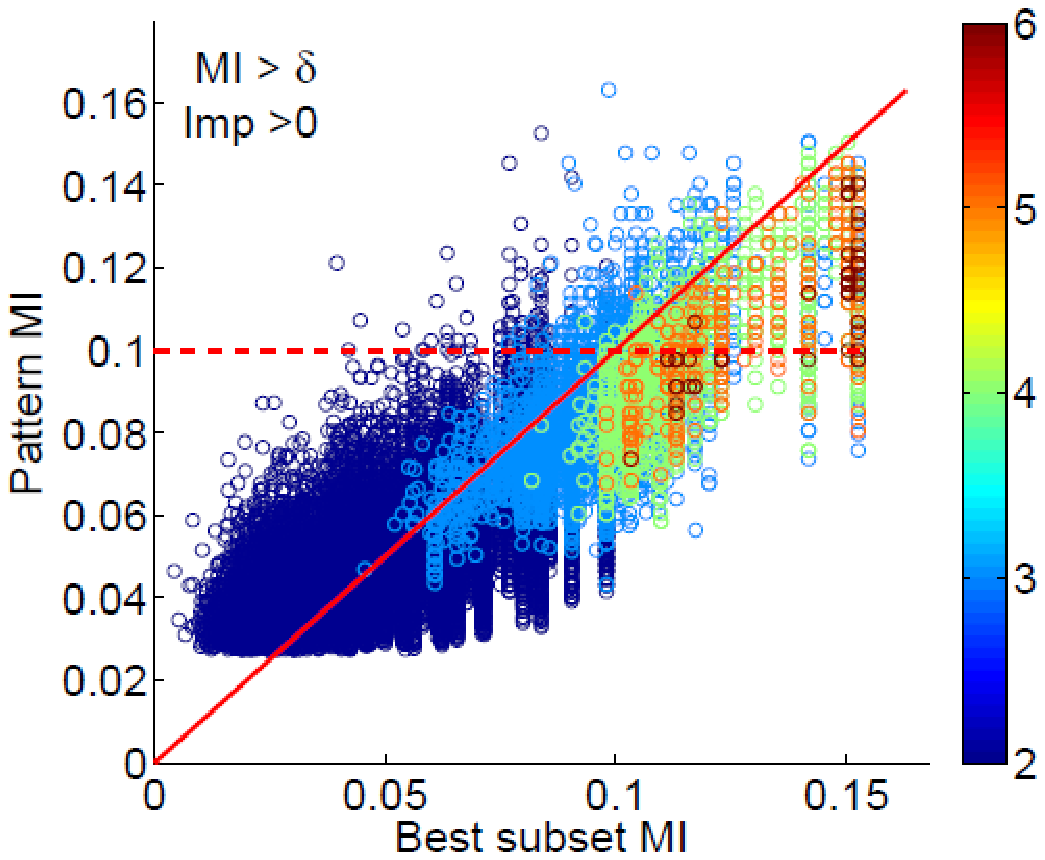}}
\subfigure[\scriptsize Positive $Syn$ (above $y=x$) \label{fig:t4inter_pre}]{\includegraphics[width=.49\textwidth]{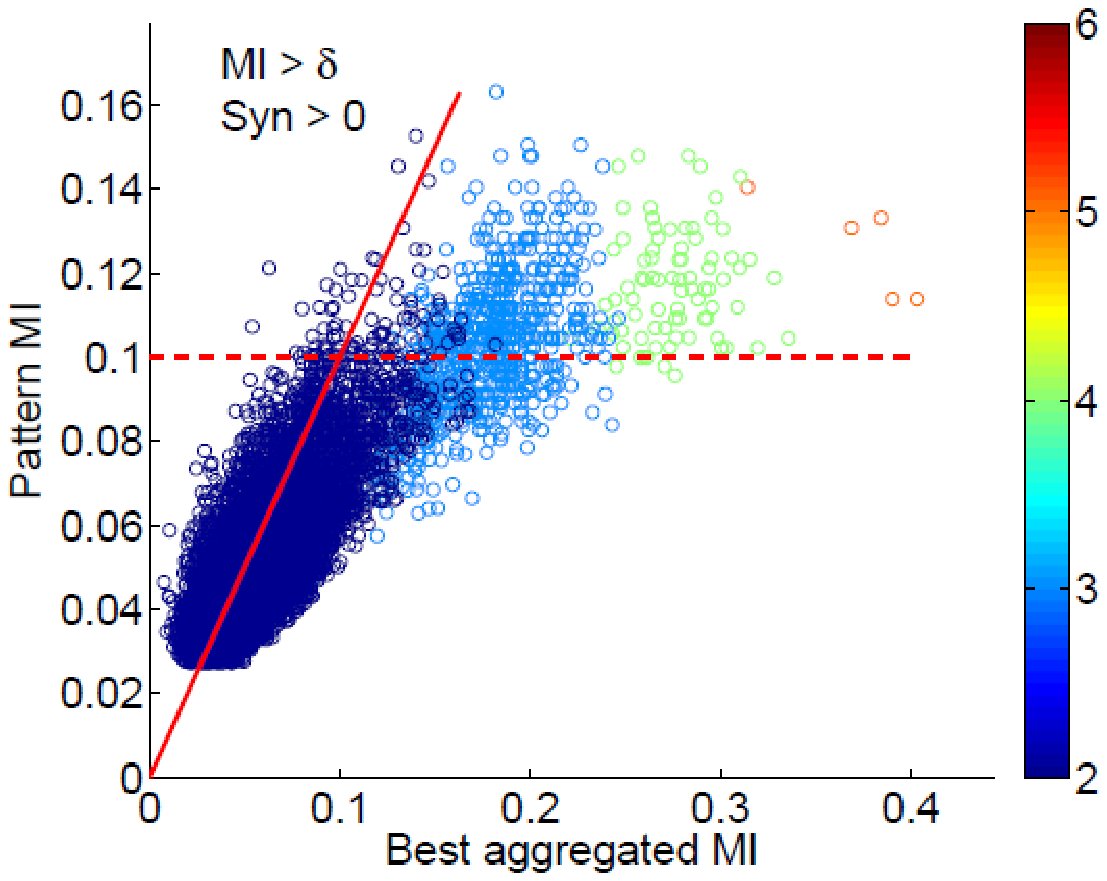}}
\caption{\small Illustration of general improvement and synergistic interaction beyond independent additive effect on the gene expression dataset. \emph{Best aggregated MI} is only computed for those patterns that have positive $Imp$.}
\label{fig:t3t4tild}
%\vspace{-0.4in}
\end{figure}

Given the definition of \emph{improvement} and \emph{synergy}, it is easy to notice that \emph{synergy} is guaranteed to be larger than \emph{improvement} (follows from the fact that MI is non-negative. Proof omitted). Essentially, \emph{synergy} is a more restrictive measure specifically for capturing interaction beyond independent addition. Figure \ref{fig:t3t4tild} compares how \emph{improvement} and \emph{synergy} capture the interaction of discriminative patterns discovered from a gene expression dataset (described in section \ref{subsec:expresults}). Figure \ref{fig:t3inter_pre} shows the $MI$ and best subset $MI$ of the discriminative patterns as a scatter plot. The horizontal dashed line indicates the cutoff values for $MI$, and the other dashed line (representing $y = x$) separate the patterns with $MI$ higher than best subset $MI$ (positive \emph{improvement}) with those that have negative \emph{improvement}. As shown, there are quite a few patterns above both the horizontal line and $y = x$, with size ranging from $2$ to $5$. In contrast to Figure \ref{fig:t3inter_pre}, the x-axis in Figure \ref{fig:t4inter_pre} is \emph{best aggregated MI} instead of \emph{best subset MI}. Corresponding to this difference, there are far fewer discriminative patterns (all of size-$2$) that are above both the horizontal line (high discriminative power) and $y = x$ (positive synergy) at the same time. This contrast is as expected given our discussion above that \emph{synergy} is a more restrictive type of interaction beyond the independent additive effect, and is guaranteed to be no more than \emph{improvement} for any pattern.

\subsubsection{Defining two different types of discriminative patterns with positive improvement}

With \emph{synergy}, we can divide all discriminative patterns with positive \emph{improvement} into two groups., i.e. those that have negative \emph{synergy} and those that have positive \emph{synergy}. Alternatively, the two groups can also be defined as those patterns that have positive \emph{improvement} (including both positive and negative \emph{synergy}) and those that specifically have positive \emph{synergy}. We take the latter route, given its simplicity in term of the definitions as shown below. Note that the observations made from both routes are essentially the same.

%\begin{definition}
%\label{def:t3inter}
\textbf{Definition 3}: An itemset $\alpha$ is a T3 discriminative pattern if the following criteria are met together for $\delta > 0$, $j > 0$:
\begin{equation}\label{equ:t3inter}
\centering
\begin{array}{ll}
(a) & MI(\alpha,C) > \delta,\\
(b) & Imp^C(\alpha) >  j.
\end{array}
\end{equation}
%\end{definition}

%\begin{definition}
%\label{def:t4inter}
\textbf{Definition 4}: An itemset $\alpha$ is a T4 discriminative pattern if the following criteria are met together for $\delta > 0$, $j > 0$:
\begin{equation}\label{equ:t4inter}
\centering
\begin{array}{ll}
(a) & MI(\alpha,C) > \delta,\\
(b) & Syn^C(\alpha) >  j.
\end{array}
\end{equation}
%\end{definition}

For illustration, we note that pattern $A$ in Figure \ref{fig:toll} is a $T4$ discriminative pattern, with $MI(A) = 0.39$, individual item $MI's$ 0.007, 0.008, and 0.029, respectively. The $M$-improvement is 0.18, while the synergy is 0.17.  

If a dataset has many $T3$ discriminative patterns, we can speculate that it contains features that complement each other for higher discriminative power in their association with the class variable. Further, if there are also many $T4$ discriminative patterns, it is expected that some features have synergistic cooperative effect beyond independent addition. In contrast, if a dataset has few or no $T3$ discriminative patterns, the discriminative features, if they exist in the dataset, are expected to be either correlated with each other ($T2$) or not form high-order combinations at all, i.e., have a very low joint frequency to pass the support threshold.

\begin{figure}[t!]
\centering
\subfigure[\scriptsize T3 Patterns (upper-right). \label{fig:t3inter_def}]{\includegraphics[width=.49\textwidth]{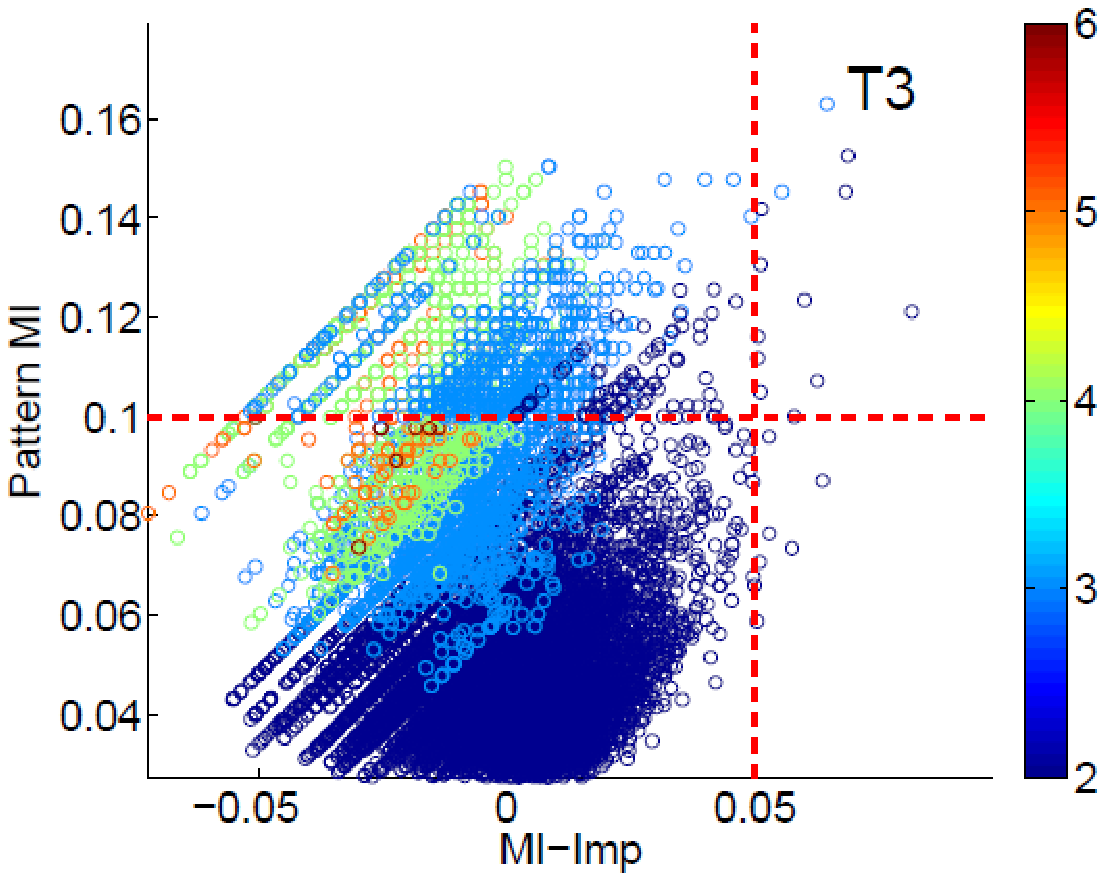}}
%\subfigure[\scriptsize A T3 example. \label{fig:t3inter_def_exp}]{\includegraphics[width=.35\linewidth]{data_T3plot_survival_101.eps}}
\subfigure[\scriptsize T4 Patterns (upper-right). \label{fig:t4inter_def}]{\includegraphics[width=.49\textwidth]{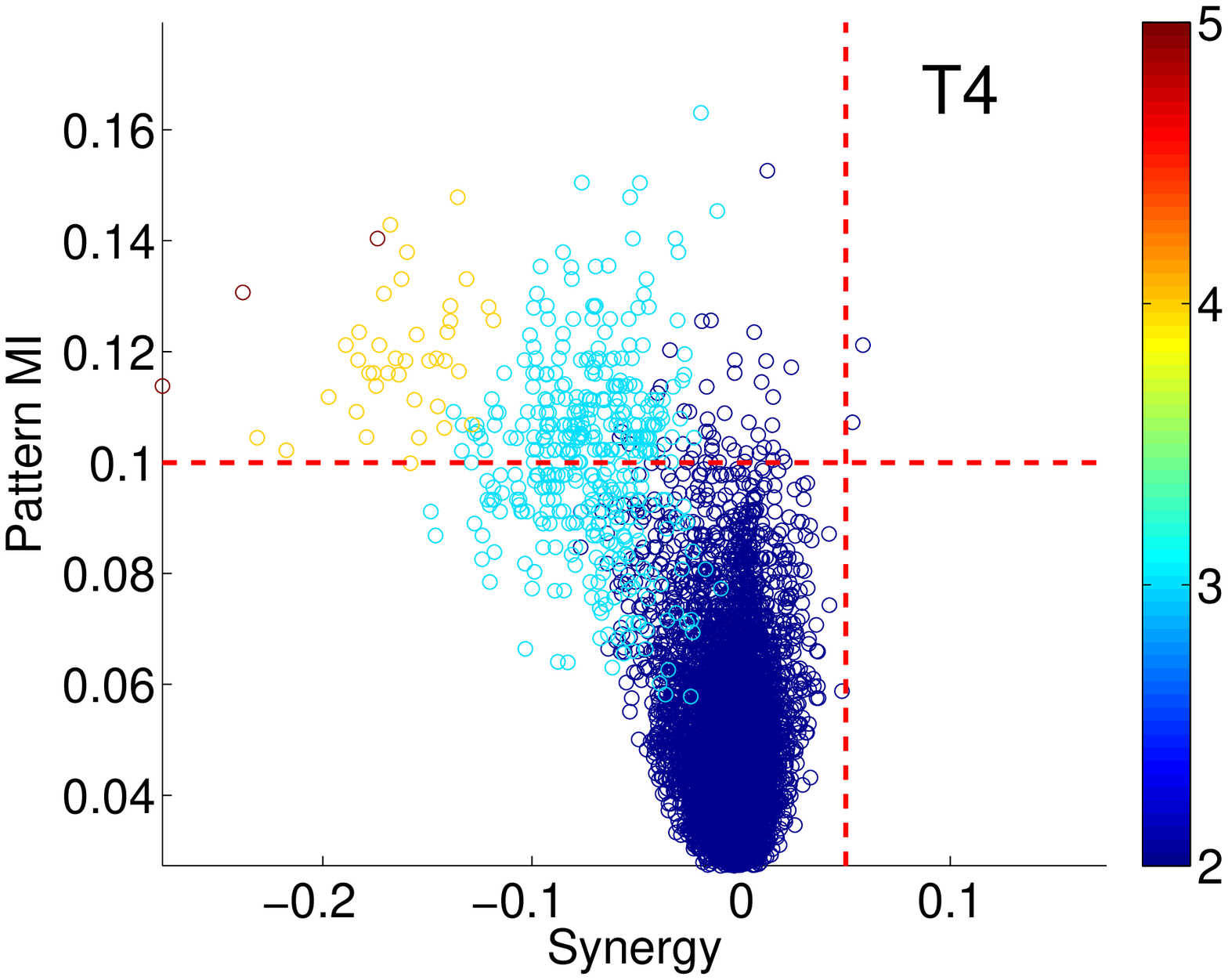}}
%\subfigure[\scriptsize A T4 example. \label{fig:t4inter_def_exp}]{\includegraphics[width=.35\linewidth]{data_T4plot_survival_101.eps}}
\caption{\small Illustration of T3 and T4 discriminative patterns on the gene expression dataset. $Syn$ is only computed for those patterns that have positive $Imp$.}
\label{fig:t3t4grid}
%\vspace{-0.4in}
\end{figure}

Figure \ref{fig:t3t4grid} shows the two sets of patterns: T3 (upper right region in Figure \ref{fig:t3inter_def}) and T4 (upper right region in Figure \ref{fig:t4inter_def}) respectively, both with $\delta = 0.1$ and $j = 0.05$. Note that, there are only two patterns (size-$2$) that have \emph{synergy} greater than $j = 0.05$. This again indicates that the synergistic interaction in T4 patterns is rare. However, as will be shown in section \ref{subsec:expresults}, these two $T4$ patterns (even very rare) are statistically significant after correcting for multiple hypothesis testing to control type I error (false discover rate $< 0.01$), and thus can be of significant interest in the biomedical domain. After all, $j = 0.05$ is an arbitrary threshold that is used to illustrate the concept. In fact, there are many other discriminative patterns with positive synergy (even though they are below $0.05$) as shown in Figure \ref{fig:t4inter_def}, which may also be interesting to specific domains. %In section \ref{subsec:expresults}, we will also show that another two there are

\begin{figure}[t!]
\centering
\subfigure[\scriptsize A T3 example.\label{fig:t3inter_exp}]{\includegraphics[width=.45\textwidth]{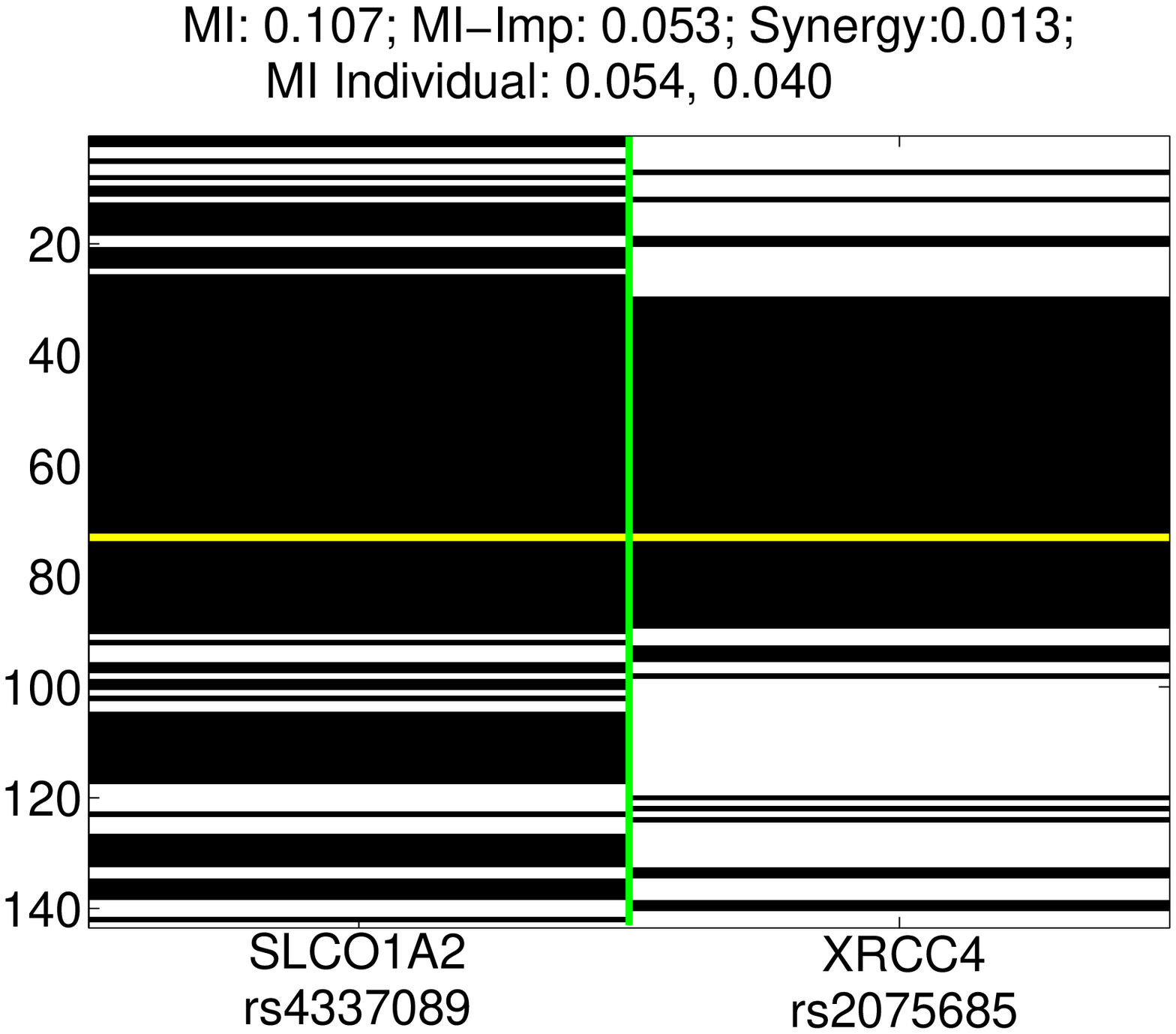}}
%\subfigure[\scriptsize A T3 example. \label{fig:t3inter_def_exp}]{\includegraphics[width=.35\linewidth]{data_T3plot_survival_101.eps}}
\subfigure[\scriptsize A T4 example.\label{fig:t4inter_exp}]{\includegraphics[width=.45\textwidth]{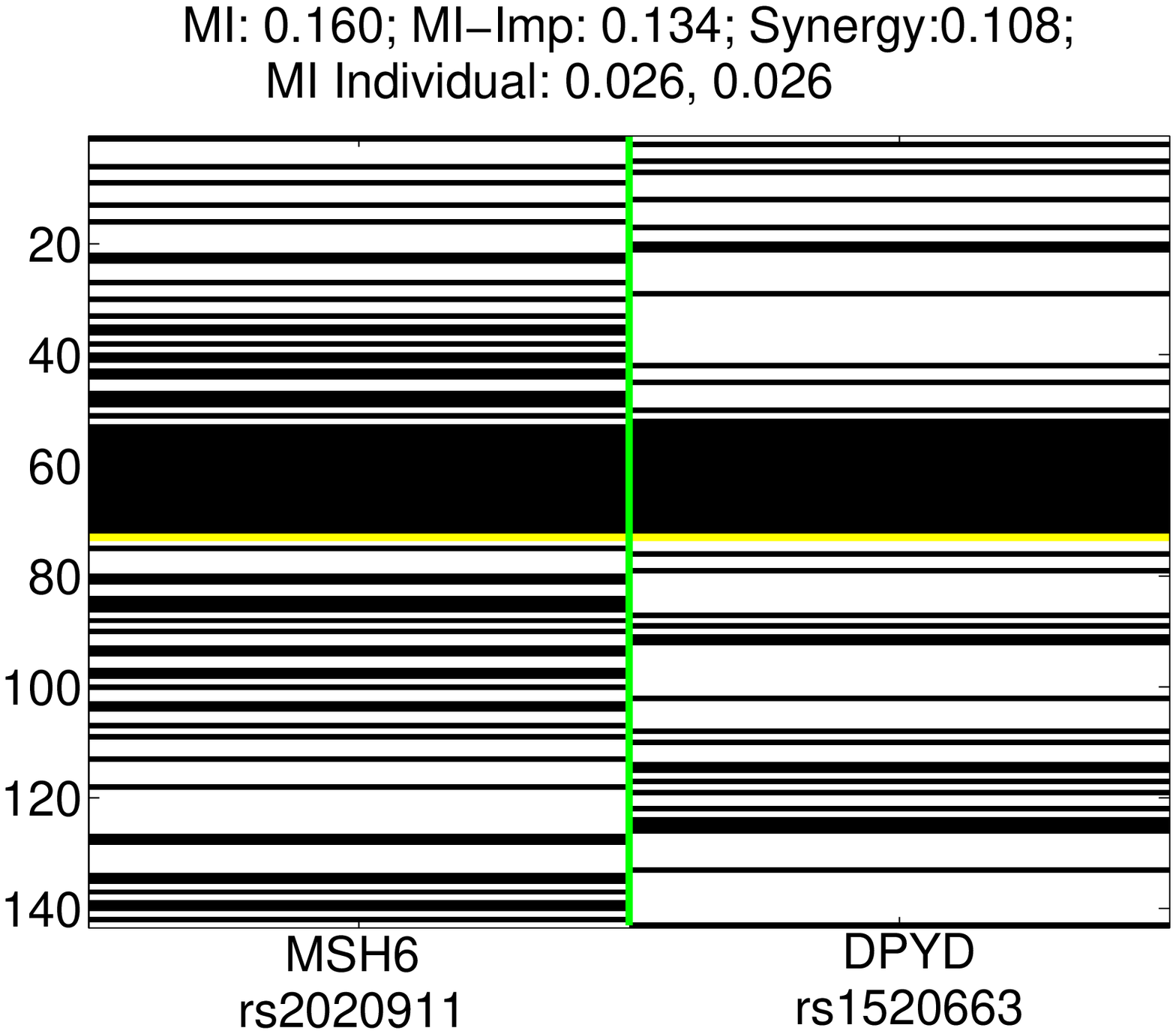}}
%\subfigure[\scriptsize A T4 example. \label{fig:t4inter_def_exp}]{\includegraphics[width=.35\linewidth]{data_T4plot_survival_101.eps}}
\caption{\small A T3 example and a T4 example, both discovered from the M-Survival SNP dataset as described in section \ref{sec:datasets} (refer to Fig. \ref{fig:gustavoPatt} for similar description.}
\label{fig:t3t4exp}
%\vspace{-0.4in}
\end{figure}

Figure \ref{fig:t3t4exp} illustrates two example patterns for $T3$ and $T4$ respectively. In Figure \ref{fig:t3inter_exp}, the individual $MIs$ of the two SNPs are $0.054$ and $0.04$ respectively. As a combination, it has a $MI$ of $0.107$, which is almost the same as the sum of the two individual $MIs$ (a low \emph{synergy} of $0.013$), indicating a independent additive effect and thus a $T3$ pattern. In contrast, the two SNPs in Figure \ref{fig:t4inter_exp} have a high \emph{synergy} of 0.108, indicating a large cooperative effect beyond independent addition. Indeed, the two genes that the two SNPs are located on, $MSH6$ and $DPYD$ are known to code proteins that have the following functions\footnote{www.genecards.org}: (i) recognizing mismatched nucleotides and (ii) catabolizing two specific types of nucleotides (uracil and thymidine), respectively. The fact that they have a synergistic interaction agrees with their closely related functions and potential compensation for each other as illustrated in Figure \ref{fig:gi_motivation_costanzo}.

\subsection{The relationships among the four different types of interactions}

In this subsection, we discuss the relationships among different types of interactions and relate other types of interactions to the four defined interactions in order to have a systematic understanding about item interactions in discriminative patterns.

\begin{figure}%[t]
	\centering
		\includegraphics[width=0.50\textwidth]{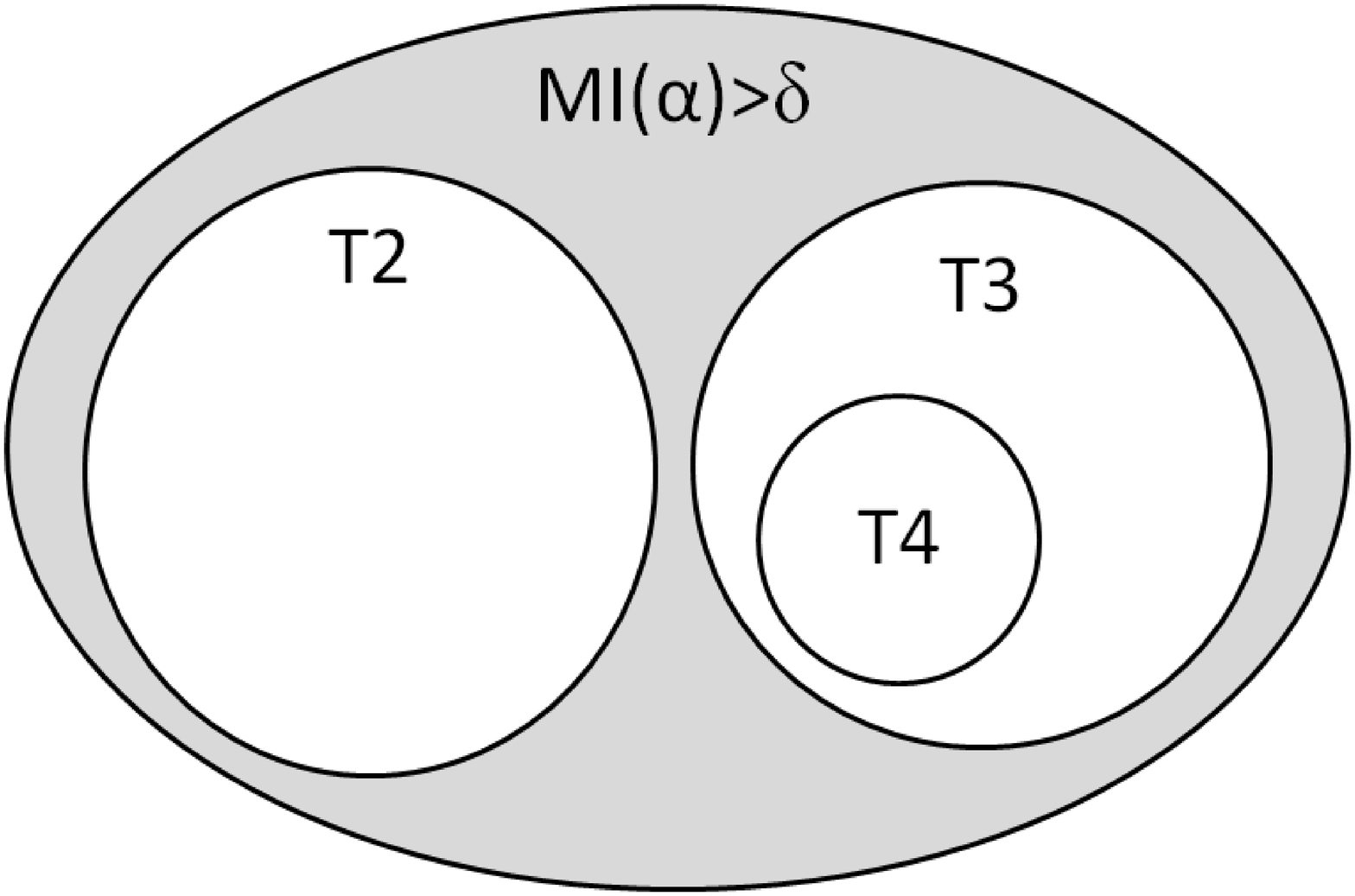}
	\caption{\small Relationship among the three types of interesting patterns for the same value of $\alpha$ and $j$.}
	\label{fig:vennConcept}
\end{figure}

Figure \ref{fig:vennConcept} shows the three interesting types of discriminative patterns (T2, T3 and T4) in the context of all discriminative patterns using a Venn diagram. The outermost circle contains all the discriminative patterns with $MI > \delta$. The set of $T3$ discriminative patterns is a superset of the set of $T4$ patterns based on the Definitions 3 and 4 (with the same $\alpha$ and $j$) and the fact that \emph{synergy} is always no more than \emph{improvement} for any pattern. The set of $T2$ discriminative patterns is disjoint with the set of $T3$ patterns, when the same value of $j$ is used in Definitions 2 and 3. Specifically, for any given value of $j$, criterion (b) in Definition 2 and criterion (b) in Definition 3 can not be met at the same time.

The next natural question is the nature of the discriminative patterns that are not any of the three types (T2, T3 and T4), i.e. the region represented by the gray background color. Indeed, they can all be considered to be in one of the two possible cases: either (i) T1 patterns with the driver-passenger interaction or (ii) the patterns, each of which can be considered as a combination\footnote{For example, if $R$ is a T2 pattern and $Q$ is a T3 pattern, then $R \cup Q$ is a combiantion of T2 and T3 pattern.} of T2 and T3 patterns. Due to the limit of space, the prove of this is available on the paper website.%, for a general discriminative pattern $\alpha$ with $MI(\alpha,C)>\delta$ and with the assumption that that $\alpha$ is neither $T2$ nor $T3$.

Note that, the goal of characterizing discriminative patterns with different types of interactions is to identify different types of interesting discriminative patterns, which are specifically $T2-T4$ in the context of this paper. It is worth noting that we do not exclude the possibility that the patterns in the gray region ($T1$ patterns, or combinations of $T2$ and $T3$ patterns) may also be interesting in some specific domains even though they are not considered as such in this paper. Thus the focus of this paper is to initiate a study of the item interactions in discriminative patterns, rather than identifying the all possible types of interesting item interactions in discriminative patterns.%First of all, all the non-interesting patterns with the driver-passenger type ($T1$) are all in there. However, in addition, there are other types of interactions that are none of the four types. The reason is, a discriminative pattern can also be a combinations of multiple patterns that are of different type ($T1~4$) after which it will no longer belong to any of the four interactions. For example, if $\alpha$ the union of two patterns ($\alpha_1$ and $\alpha_2$) with $T2$ and $T3$ interactions respectively and if $A$ and $B$ share the same supporting transactions (i.e. $MI(\alpha) > \delta$), then $\alpha$ is none of the four types of patterns: (i) it is not $T1$This indicates that the items in $\alpha$ are not all consistent drivers, and thus there are passengers (in a general sense rather than the specific definition in criterion (c) if Definition \ref{def:t1inter}) in $\alpha$. Thus, $\alpha$ can be considered as a $T1$ pattern. 

\subsection{Correction for Multiple Hypothesis Tests}
\label{sec:interFDR}

As discussed by recent work \cite{webb2007discovering,gionis2007assessing,heikki2009tellme}, an association pattern mining task (e.g. frequent patterns, discriminative patterns) essentially conducts a large number of hypothesis testing. Thus, in order to control type I error (due to the multiple hypothesis testing), corrections on the significance of the discovered patterns is necessary. Among different approaches for correcting multiple hypothesis testing, the randomization based approaches \cite{gionis2007assessing} are non-parametric and thus more reliable in term of not introducing bias. Randomization frameworks have been extensively explored in the context of frequent pattern mining and clustering \cite{heikki2009tellme}. For discriminative pattern mining, a special type of randomization procedure is needed, in which the randomization is performed by shuffling the class labels for the samples. For the details of the randomization and the calculation of corrected p-value or false discovery rate (FDR), refer to \cite{fang2010tkde,fang2010subspace,subramanian2005gsa}. In section \ref{subsec:expresults}, we will show that many of the discovered $T2-T4$ patterns are statistically significant after correcting for multiple hypothesis tests. %After all, we want to note that in practice, the characterization of discriminative patterns can be done either directly on the set of discovered patterns or after the correction of multiple hypothesis testing is done. %Due to the space limit, we will focus on the characterization of different types of interactions without correcting the multiple hypothesis testing in the experiment section.

 %This section is just a place holder and it is not needed in this paper because the focus of the paper is on different types of interactions, which is orthogonal with the statistical significance of discriminative patterns. In conclusion and future work, there is a sentence saying that after grouping these different types of discriminative patterns, randomization tests can be used to correct for type I error (false positives).

\section{Experiments}
\label{sec:interresults}

In this section, we use a variety of real datasets to demonstrate the existence, properties and statistical significance of different types of discriminative patterns that we characterized in section \ref{sec:intertypes}. We also show how the characterization can provide novel insights into discriminative pattern mining and the discriminative pattern structure of different datasets, beyond those provided by current approaches that focus mostly on pattern-based classification and subgroup discovery.

\subsection{Data Sets}
\label{sec:datasets}

We use the following three different types of real datasets, with details summarized in Table \ref{table:datasets} and detailed pre-processing steps described on the paper website:

\textbf{Ten UCI datasets} \cite{uci2007} with a variety of dimensions and densities (Table \ref{table:datasets}).

\textbf{A gene expression dataset} on breast cancer \cite{VijverData2002} (pre-processed as suggested in \cite{veer02} and binarized as done in \cite{fang2010tkde,sanjay2007local}. We denote this dataset by Breast(GEP).

\textbf{Two single-nucleotide polymorphism (SNP) datasets}: SNP profile captures the genetic variations of a person at single-nucleotide resolution, which are commonly used in disease-association studies \cite{multigene2004nature,snpPair2007science, snpPair2007naturegene}. The diseases studied with these two datasets are Myeloma \cite{brian2008bmc} and lung cancer \cite{church2009prospectively} respectively. We denote these two datasets by M-Survival(SNP) and Lung(SNP).

\begin{table}
\centering
\scriptsize
\begin{tabular}{|c|c|c|c|c|c|c|c|c|c|}
\hline
Datasets & Size of & Size of & \# of & Density & $MI>\delta$ & \# of T2 & \# of T3 & \# of T4 & \% of $T2-T4$\\
& ($+$ class) & ($-$ class) & Items & & & & & & patterns\\ \hline
Breast (GEP) & 217 & 78 & 11962 & 0.1662 & 1642 & 240 (29)& 13 (21) & 2 (4) &0.154 \\ \hline
Lung (SNP) & 96 & 99 & 8777 & 0.3855 & 6 & 0 & 4 (4) & 2 (3)& 0.667 \\ \hline
M-Survival (SNP ) & 70 & 73 & 8265 & 0.3325 & 62 & 0 & 32 (42) & 16 (27)& 0.516 \\ \hline
Chess (UCI) & 1527 & 1669 & 73 & 0.4932 & 109 & 0 & 5(7) & 1(3) & 0.046 \\ \hline
Sonar (UCI) & 111 & 97 & 42 & 0.5 & 19476 & 268 (37) & 21 (17) & 0 & 0.015\\ \hline
Hepatic (UCI) & 32 & 123 & 33 & 0.4561 & 3144 & 16 (11) & 6 (7)& 0 & 0.007 \\ \hline
Cleve (UCI) & 165 & 138 & 27 & 0.4074 & 708 & 21 (12) & 2 (4) & 0 & 0.032 \\ \hline
Horse (UCI) & 232 & 136 & 57 & 0.2234 & 106 & 1 (2) & 1 (2) & 0 & 0.019 \\ \hline
Adult (UCI) & 11687 & 37155 & 94 & 0.1371 & 661 & 4 (6) & 0 & 0 & 0.006\\ \hline
Crx (UCI) & 307 & 383 & 50 & 0.2784 & 625 & 3 (6) & 0 & 0 & 0.005 \\ \hline
Hypo (UCI) & 151 & 3012 & 50 & 0.4524 & 644 & 1 (2) & 0 & 0 & 0.002 \\ \hline
Mushroom (UCI) & 3916 & 4208 & 118 & 0.1923 & 2334 & 0 & 0 & 0 & 0 \\ \hline
Waveform (UCI) & 1657 & 1647 & 102 & 0.1863 & 17 & 0 & 0 & 0 & 0 \\ \hline
\end{tabular}
\caption{\scriptsize Details of each dataset and a summary of the number of different types of discriminative patterns discovered. For column $7-9$, in addition to the number of discovered patterns, we also show (in the bracket) the number of unique items in the union of the set of patterns to reflect the redundancy among the patterns.  $\delta = 0.1$ and $j= 0.05$ are used for all the datasets.}
\label{table:datasets}
\end{table}

\subsection{Experimental Results}
\label{subsec:expresults}

For each dataset, we first discover a set of discriminative patterns with existing algorithms. Specifically, for the dense and high dimensional datasets (Breast (GEP), the two SNP datasets, Chess (UCI) and Hypo (UCI)), we leverage the SMP algorithm proposed in \cite{fang2010tkde} to discover discriminative patterns ($SupMaxPair = 0.2$). For the other sparse or low-dimensional datasets we simply use FPC \cite{fpclose2003} with $minsup = 10\%$, because SMP may miss some high-support patterns although it is efficient on discovering discriminative patterns from dense and high-dimensional data \cite{fang2010tkde}.

\begin{figure}[t!]
\centering
%\subfigure[\scriptsize Support ratio vs. Maximal-subset support ratio. \label{fig:mVSmax_rs}]{\includegraphics[width=.45\linewidth]{datauci_hepatic_plot1_101.eps}}
%\subfigure[\scriptsize Breast(GEP) T2 \label{fig:survivalt2}]{\includegraphics[width=.32\linewidth]{.eps}}
%\subfigure[\scriptsize Breast(GEP) T3 \label{fig:survivalt3}]{\includegraphics[width=.32\linewidth]{.eps}}
%\subfigure[\scriptsize Breast(GEP) T4 \label{fig:survivalt4}]{\includegraphics[width=.32\linewidth]{.eps}}
%\\
\subfigure[\scriptsize M-Survival(SNP) T2 \label{fig:survivalt2}]{\includegraphics[width=.27\textwidth]{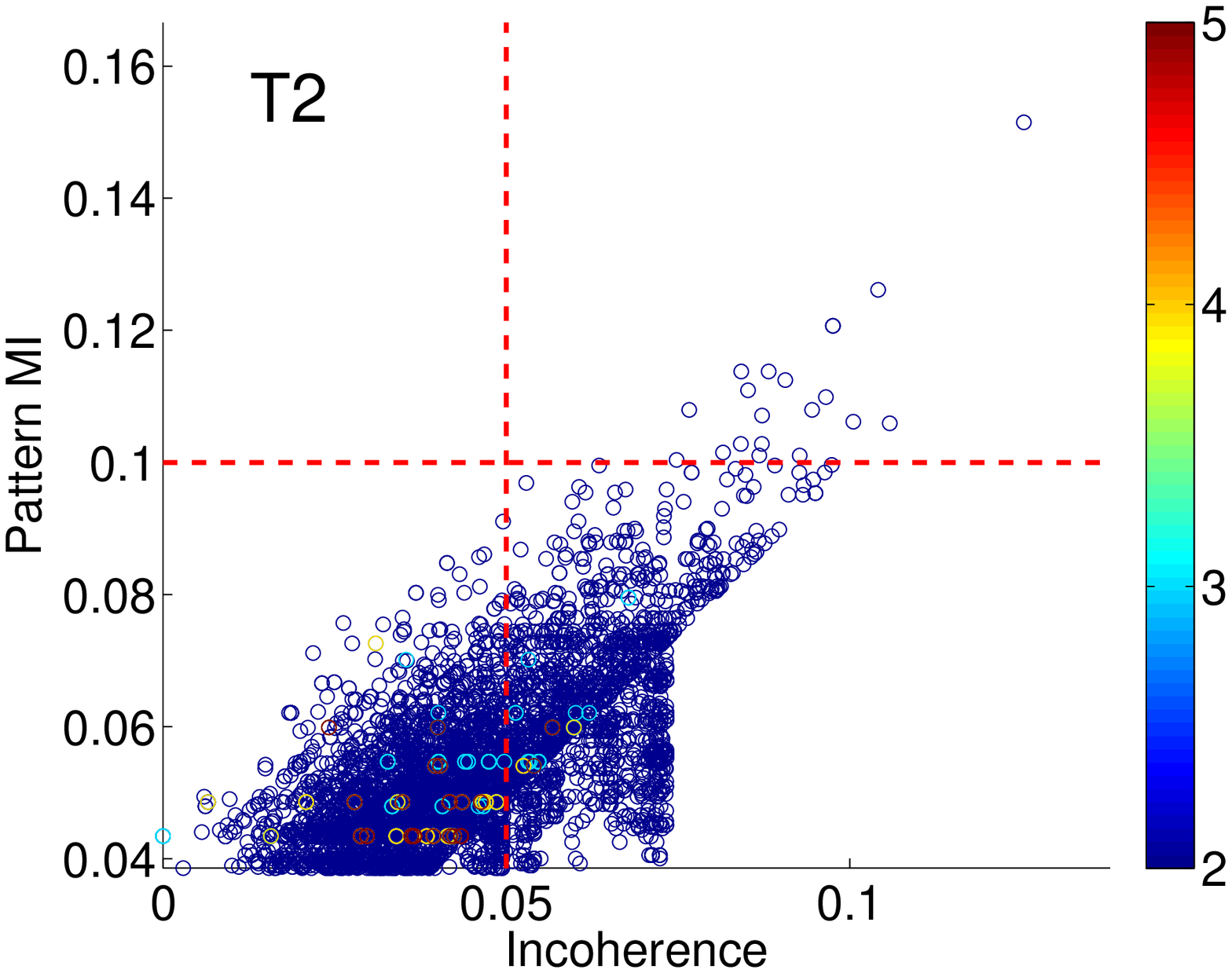}}
\subfigure[\scriptsize M-Survival(SNP) T3 \label{fig:survivalt3}]{\includegraphics[width=.27\textwidth]{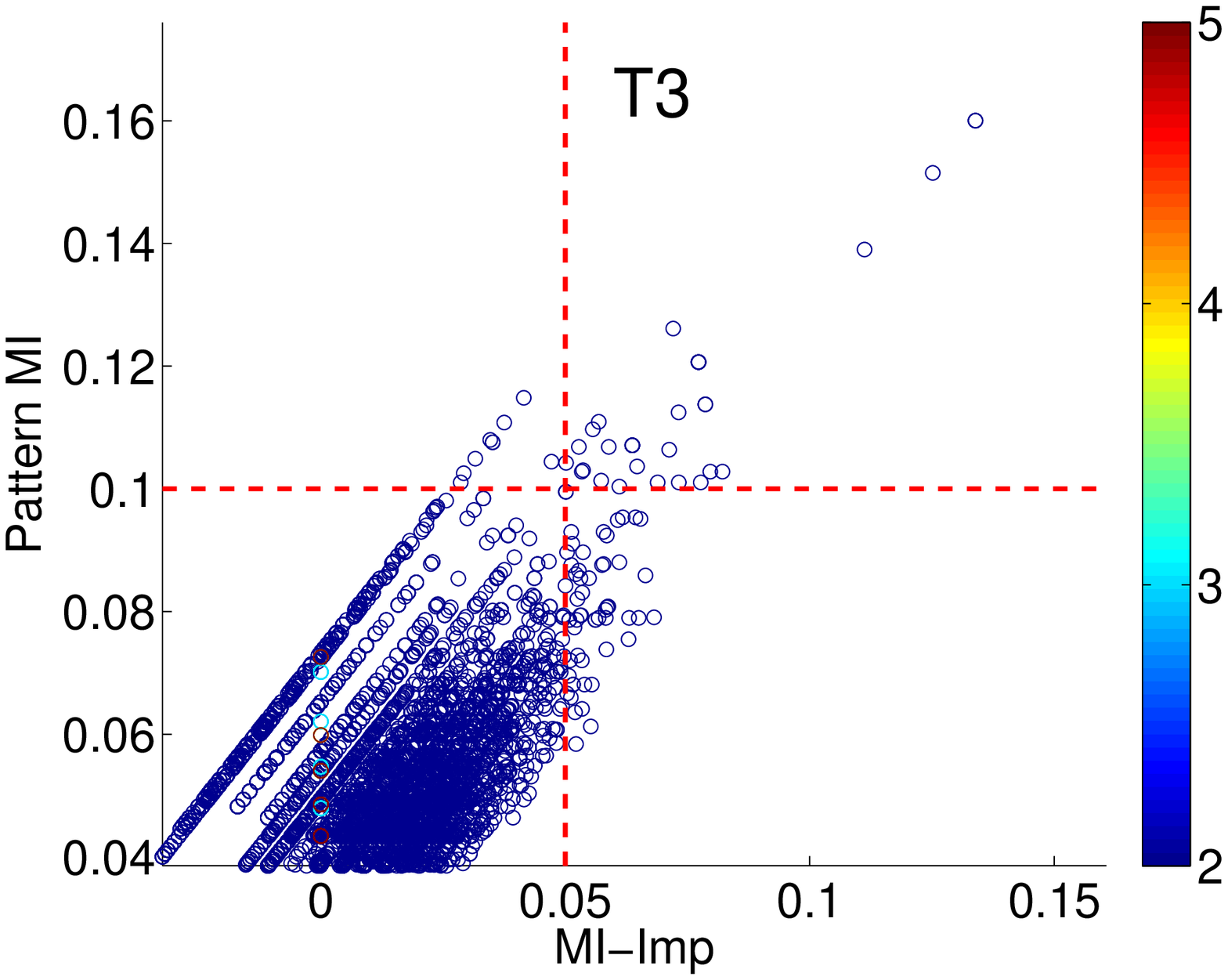}}
\subfigure[\scriptsize M-Survival(SNP) T4 \label{fig:survivalt4}]{\includegraphics[width=.27\textwidth]{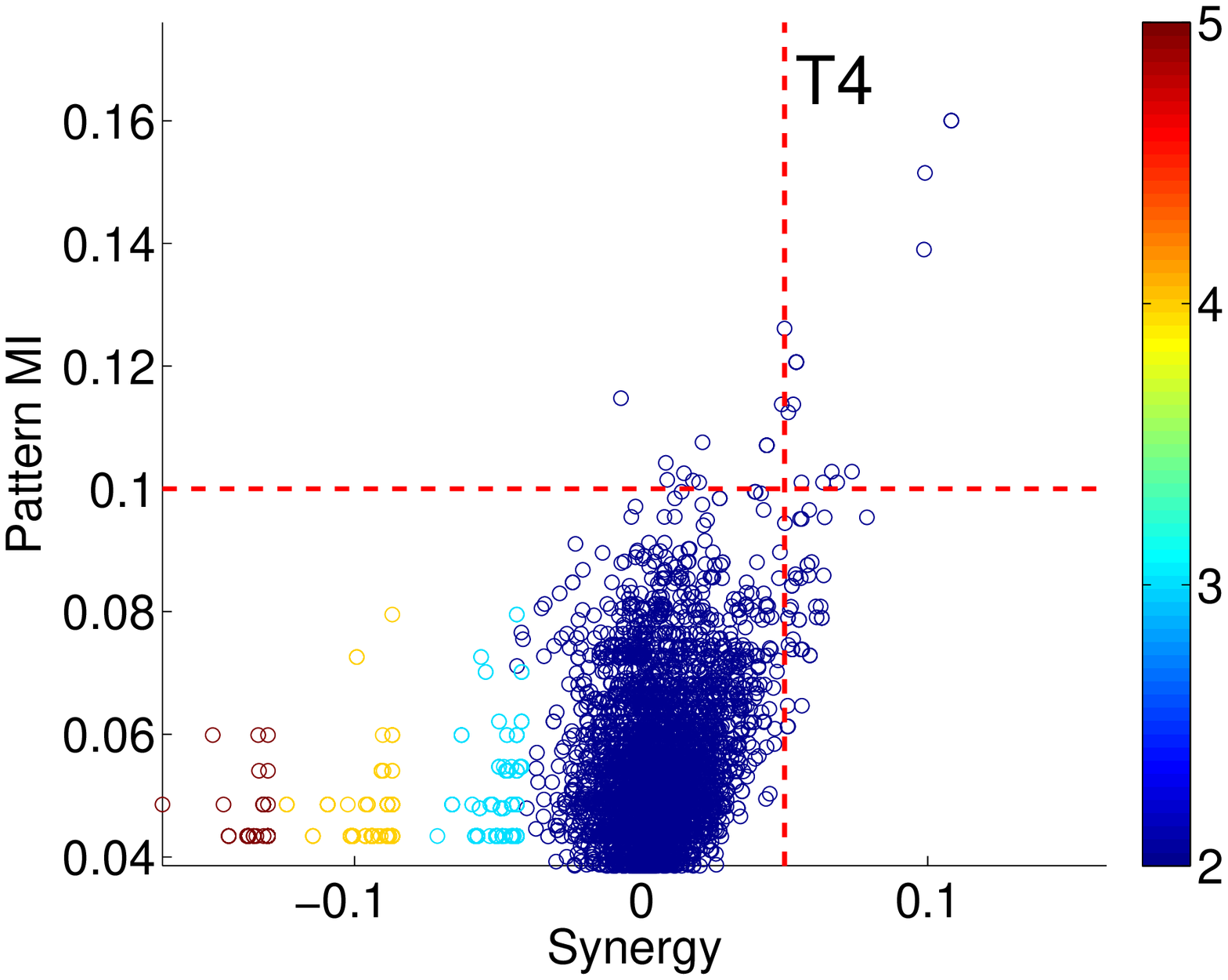}}
%\\
%\subfigure[\scriptsize Lung (SNP) T2 \label{fig:lungt2}]{\includegraphics[width=.32\linewidth]{dataSNP_lungcancer_plot4_101.eps}}
%\subfigure[\scriptsize Lung (SNP) T3 \label{fig:lungt3}]{\includegraphics[width=.32\linewidth]{dataSNP_lungcancer_plot4_102.eps}}
%\subfigure[\scriptsize Lung (SNP) T4 \label{fig:lungt4}]{\includegraphics[width=.32\linewidth]{dataSNP_lungcancer_plot4_103.eps}}
\\
\subfigure[\scriptsize Sonar(UCI) T2 \label{fig:sonart2}]{\includegraphics[width=.27\textwidth]{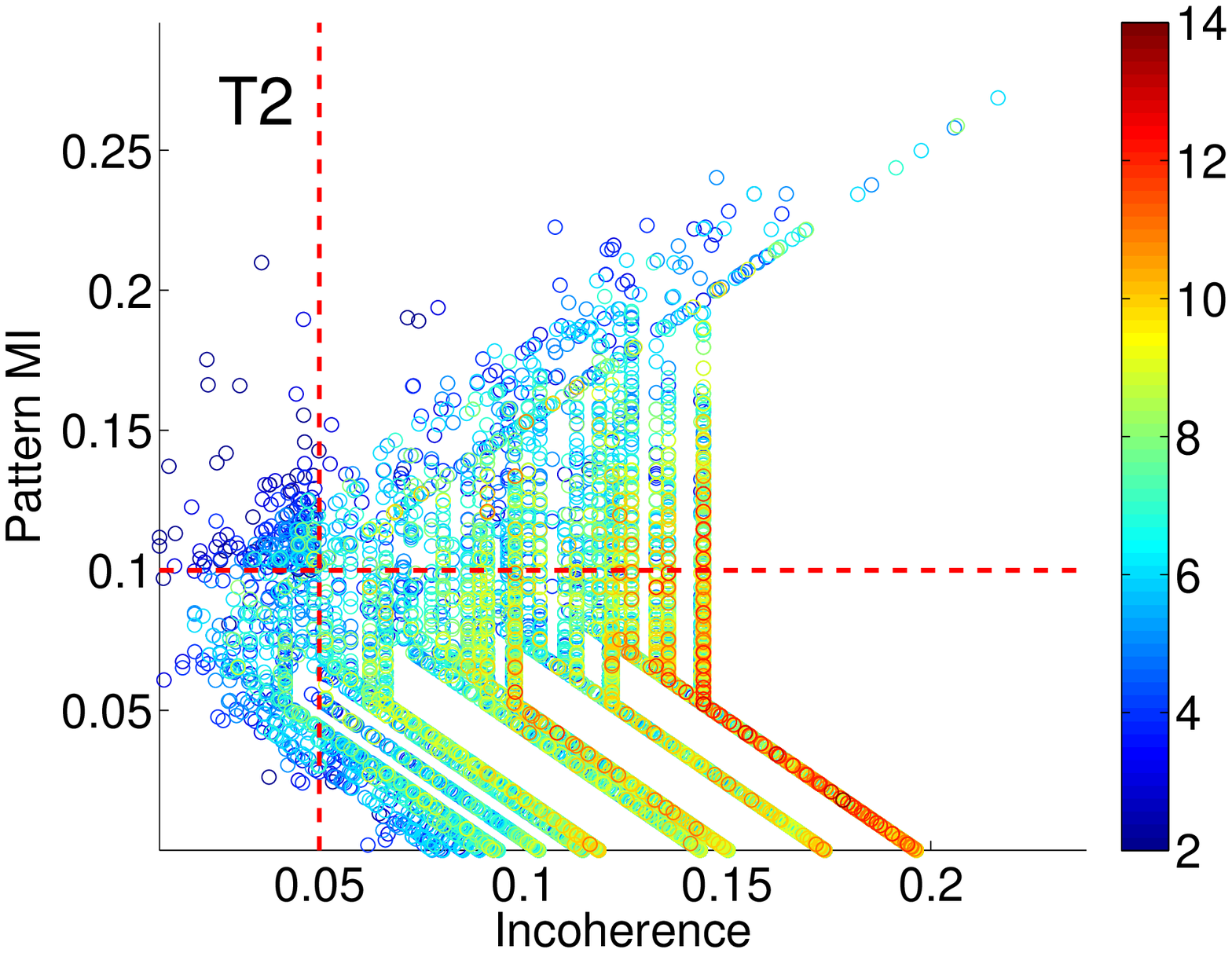}}
\subfigure[\scriptsize Sonar(UCI) T3 \label{fig:sonart3}]{\includegraphics[width=.27\textwidth]{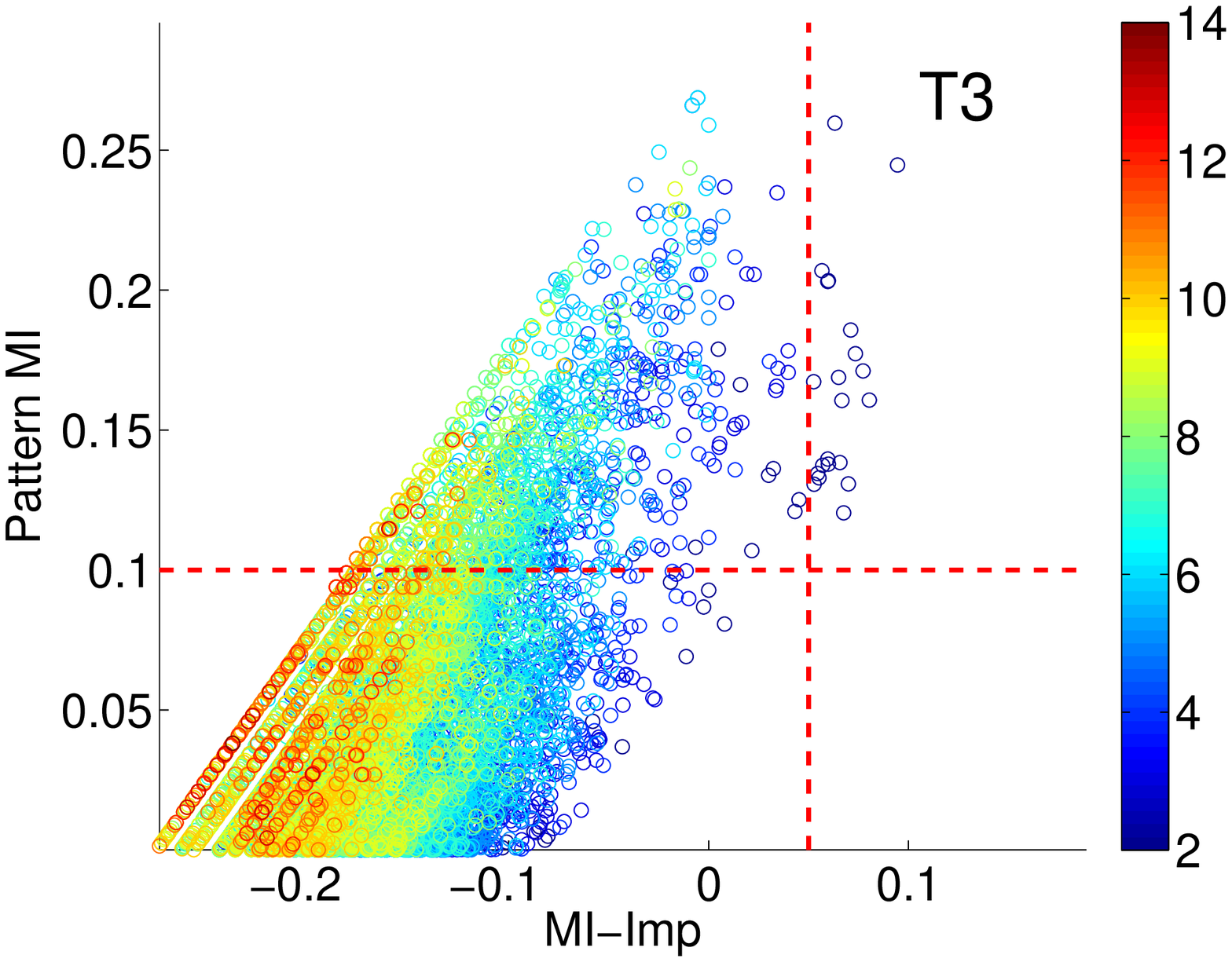}}
\subfigure[\scriptsize Sonar(UCI) T4 \label{fig:sonart4}]{\includegraphics[width=.27\textwidth]{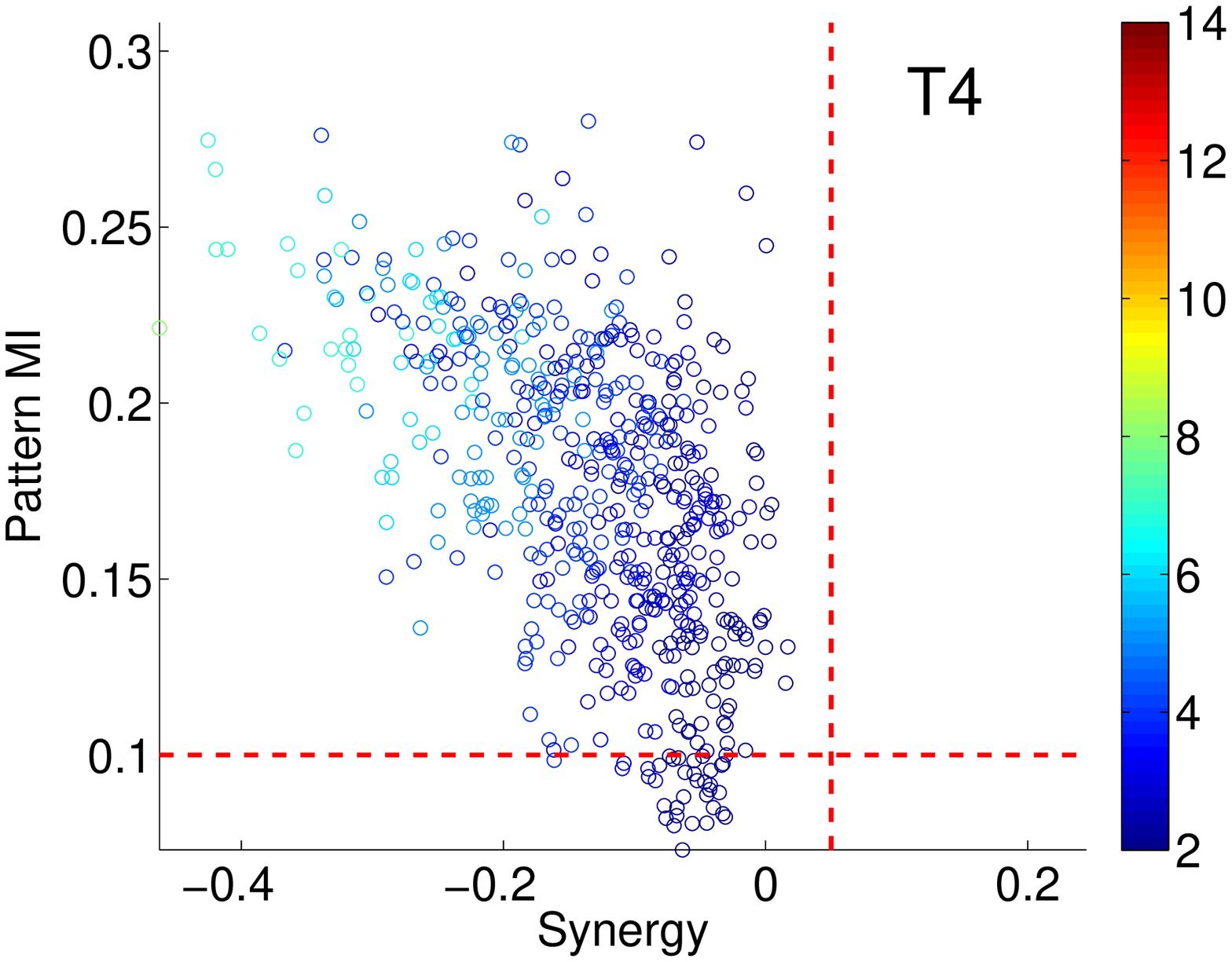}}
%\\
%\subfigure[\scriptsize Hepatic(UCI) T2 \label{fig:Hepatict2}]{\includegraphics[width=.27\linewidth]{datauci_hepatic_plot4_101.eps}}
%\subfigure[\scriptsize Hepatic(UCI) T3 \label{fig:Hepatict3}]{\includegraphics[width=.27\linewidth]{datauci_hepatic_plot4_102.eps}}
%\subfigure[\scriptsize Hepatic(UCI) T4 \label{fig:Hepatict4}]{\includegraphics[width=.27\linewidth]{datauci_hepatic_plot4_103.eps}}
%\\
%\subfigure[\scriptsize Cleve(UCI) T2 \label{fig:Clevet2}]{\includegraphics[width=.27\linewidth]{datauci_cleve_plot4_101.eps}}
%\subfigure[\scriptsize Cleve(UCI) T3 \label{fig:Clevet3}]{\includegraphics[width=.27\linewidth]{datauci_cleve_plot4_102.eps}}
%\subfigure[\scriptsize Cleve(UCI) T4 \label{fig:Clevet4}]{\includegraphics[width=.27\linewidth]{datauci_cleve_plot4_103.eps}}
\caption{\small Existence of $T2-T4$ patterns in two representative datasets: (a)-(c) M-Survival (SNP); (d)-(f) Sonar (UCI). In subfigures (c) and (f), \emph{synergy} is only computed for those patterns that have positive \emph{improvement}.}
\label{fig:t2t3t4uci}
%\vspace{-0.4in}
\end{figure}

For each set of discovered patterns (only closed itemsets), we apply the criteria of the three types of discriminative patterns ($T2-T4$) presented in section \ref{sec:intertypes} and get the number of patterns for each type. Figures \ref{fig:t2t3t4uci} illustrate the existence of T2,  T3 and T4 discriminative patterns in the representative SNP dataset (subfigures (a)-(c)), and the representative UCI datasets (subfigures (d)-(f)). Note that, the similar set of figures for the gene expression dataset can be found in section \ref{sec:t3t4now}, i.e. Figures \ref{fig:t2onAdult}, \ref{fig:t3inter_def} and \ref{fig:t4inter_def}.

Several observations about each type of interactions can be made from Table \ref{table:datasets} and Figure \ref{fig:t2t3t4uci}.

\textbf{1: T2 discriminative patterns are common in most UCI datasets and the gene expression dataset, but not in the SNP datasets:} On one hand, this indicates that the UCI datasets and the gene expression dataset have features that are discriminative and correlated with each other. On the other hand, the fact that the SNP datasets do not have T2 patterns indicates that the discriminative SNPs are not correlated with each other. In addition, column $7$ in Table \ref{table:datasets} indicates that the across-pattern redundancy is high in $T2$ patterns, i.e. the number of unique items is generally much smaller than the number of patterns.%either have observation also implies the genetic complexity in SNP datasets, i.e. individual are not differentiating as individuals). In fact, we will show next that SNPs are generally not individually discriminative but can form $T3$ patterns. 
	
\textbf{2: T3 discriminative patterns exist in about half of the UCI datasets and all three of the biological datasets:} These datasets are expected to contain discriminative features that are complementary to each other in their improved discriminative power as a pattern. In contrast, the other datasets that have very few or no $T3$ discriminative patterns, the discriminative features, if they exist in the dataset, are expected to be either correlated with each other ($T2$) such as Cleve (UCI) or simply do not contain interesting feature combinations (independent association with the class variable) such as Mushroom and Waveform. The fact that the three biological datasets have many $T3$ discriminative patterns is consistent with common knowledge that complex diseases involve the cooperation of multiples genes. This is especially true for the the two SNP datasets, where there are no $T2$ discriminative patterns but many $T3$ discriminative interactions. In addition, column $8$ shows that the across-pattern redundancy in $T3$ patterns is lower than in $T2$ patterns, because the number of unique items is generally similar as the number of patterns.% in contrast to column $7$ for T2 patterns.%This indicates that \emph{improvement} is ubiquous in discriminative patterns .
	
\textbf{3: T4 discriminative patterns exist in all three of the biological datasets and only one UCI dataset (Sonar):} First, T4 pattern is rare because $T4$ is based on the most restrictive type of interaction (\emph{synergy}). Nevertheless, the fact that the gene expression and SNP datasets contain many $T4$ discriminative patterns indicates the relatively higher complexity in genetic datasets compared to the common UCI datasets. In addition, column $9$ shows that the across-pattern redundancy in $T4$ patterns is similar as in $T3$ patterns, which are both lower than in $T2$ patterns.
	
\textbf{4: The number of $T2-T4$ patterns is much smaller compared to the overall number of discriminative patterns:} The last column in Table \ref{table:datasets} shows the fraction of discriminative patterns that are either $T2$, $T3$ or $T4$ (the three interesting types). Except for the two SNP datasets, the fractions are generally very low, which indicate that many discriminative patterns with good discriminative power are not interesting from the perspective of the interestingness considered in this paper. The extreme cases are the Mushroom and Waveform datasets, which do not contain any of the three types of patterns. This indicates that the discriminative features are neither uncorrelated with each other nor complementary to each other in these two datasets, i.e. independently discriminative features. This observation indicates that the actual number of interesting patterns is much more manageable compared to the huge number of patterns that are generally encountered without a detailed characterization. % More words will be added (indicating only singletons).
	
\textbf{5: $T2-T4$ patterns generally have smaller size compared to the entire set of discriminative patterns:} From the color of the circles in the figures, $T2-T4$ patterns are generally of size $2-6$. This is in contrast to the wider range of sizes for the entire set of discriminative patterns, which can be as high as $14$. This agrees with the observations made in the recent work on constraint-based generation of high-order discriminative patterns \cite{mike2011pakdd}. Specifically, Steinbach et al. observed that the larger (size) an itemset becomes, the harder it is for the itemset to meet the constraints for a discriminative pattern, when the constraints are not only on the discriminative power of the pattern but also the \emph{improvement} of the discriminative power. This also suggests that the computational complexity of discriminative pattern mining could be less than expected given that too large patterns tend to be uninteresting in term of the meaningful relationships scoped in this paper.
	
\textbf{6: $T2-T4$ patterns discovered from all the datasets are statistically significant}. In the columns $7-9$ in Table \ref{table:datasets}, all the $T2-T4$ patterns are statistically significant with $FDR < 0.01$ after correcting for multiple hypothesis tests to control type I error (method discussed in section \ref{sec:interFDR}). Specifically for the three biological datasets, the characterization of those statistically significant gene or SNP combinations can assist the further biological interpretations, and reveal novel insights to the mechanisms of complex diseases.

The above comprehensive observations illustrate the existence, characteristics and statistical significance of the different types of patterns. They also illustrate how the proposed framework can provide novel insights into discriminative pattern mining and the discriminative pattern structure of different datasets. %For domains such as biomedical and genetic research, differentiating these different types of interactions is critical because they lead to different types of biological interpretations. For general application domains, this interaction-based characterization framework provides a novel and informative view about the composition of discriminative patterns in a dataset, beyond existing studies that focus mostly on pattern-based classification and subgroup discovery.

\section{Related Work}
\label{sec:interrelatedwork}

Over the past decade, many approaches have studied discriminative patterns and related topics. The most relevant related work was discussed earlier in Section \ref{sec:intro}. Among other work focusing on mining discriminative patterns, the most relevant ones are \cite{li2001jumping,loekito2006kdd_highd,li2007delta,morishita2000pod_stat}. Many existing approaches also used discriminative pattern for classification \cite{cbabin2001,hongcheng2007icde,hongcheng2008icde,hongcheng2008kdd,yan2008leap,lo2009classification}. Additional related papers in the area include \cite{li2007delta,kraljnovak2009cset,loekito2006kdd_highd,petra2008jmlr,soulet2004condensed,fang2010tkde}. We also refer the readers to a comprehensive survey on discriminative patterns by Novak et al. \cite{petra2009jmlr}. 

\section{Conclusion}
\label{sec:interconclusion}

In this paper, we categorized discriminative patterns into four groups based on item interactions: (i) driver-passenger, (ii) coherent, (iii) independent additive and (iv) synergistic beyond independent addition. The coherent, additive, and synergistic patterns are of practical importance, with the latter two representing a gain in the discriminative power of a pattern over its subsets. Synergistic patterns are most restrictive, but perhaps the most interesting since they capture a cooperative effect that is more than the sum of the effects of the individual items in the pattern. The experiments provided a number of insights into the nature of discriminative patterns in various real datasets and the characteristics of the different types of patterns.

Particularly worth noting is that all types $T2-T4$ patterns were significant in all the datasets for which we evaluated pattern significance. While this needs to be investigated further, we believe that this is mostly due to the pruning of a large number of patterns that are not likely to be of interest. Without such pruning, the number of patterns is typically very large, as is typical in most types of association analysis, and thus, the FDR of the resulting patterns tends to be low unless the patterns are very strong since FDR depends very heavily on the number of patterns being considered. We are hopeful that this observation will allow discriminative pattern mining to be more effectively used for a wide variety of applications, both in the biomedical area and beyond. 

Several further directions can be explored in the future. (1) The four types of patterns defined in the paper are mainly based on the building-block measure \emph{mutual information} to make the presentation consistent and easy to follow, and other statistical measusres can also be explored as building-block measures or specifically for a certain type of pattern. For instance, the logistic regression-based measure studied in \cite{storey2005multiple} can be leveraged as an alternative to \emph{synergy}, (2) Other types of interactions can be explored especially those that may be interesting to specific domain but are considered as non-interesting in the context of this paper. (3) From the computational perspective, it is also interesting to design mining algorithms that can directly search for a particular type of discriminative patterns, which is expected to be much faster given a more specific definition, whose anti-monotonic properties can be leveraged as additional pruning constraints.% (4) Recent work has proposed using constraints to generate synthetic discriminative patterns \cite{mike2011pakdd}, and it would be interesting to design specific constraints to generate each of the three interesting types of discriminative patterns presented in this paper. 

\begin{comment}
\end{comment}

\section{Acknowledgments}
The authors thank Hong Cheng for providing the discretized UCI datasets used in the paper.

\end{document}